\documentclass[a4paper,11pt]{article}
\pdfoutput=1 

\usepackage{jcappub} 

\usepackage{aas_macros}
\usepackage{natbib}
\usepackage{amsfonts}
\usepackage{amsmath}
\usepackage{amssymb}
\usepackage{graphicx}
\usepackage[dvipsnames]{xcolor}
\usepackage{hyperref}
\hypersetup{colorlinks=true,allcolors=teal}
\usepackage{caption}
\usepackage{subcaption}

\newcommand{\kk}{\ensuremath{\mathbf{k}}}
\newcommand{\xx}{\ensuremath{\mathbf{x}}}
\newcommand{\Msun}{\ensuremath{M_{\odot}}}

\newcommand{\Mpch}{\ensuremath{h^{-1}{\rm Mpc}}}
\newcommand{\hMpc}{\ensuremath{h\,{\rm Mpc}^{-1}}}
\newcommand{\kpch}{\ensuremath{h^{-1}{\rm kpc}}}
\newcommand{\kms}{\ensuremath{{\rm km\,s}^{-1}}}

\newcommand{\avg}[1]{\ensuremath{\left\langle \,#1\, \right\rangle}}
\newcommand{\e}[1]{\ensuremath{{\rm e}^{#1}}}

\newcommand{\eqn}[1]{equation~\eqref{#1}}

\newcommand{\be}{\begin{equation}}
\newcommand{\ee}{\end{equation}}

\usepackage{multirow}
\usepackage{tabularx}

\newcommand{\Om}{\ensuremath{\Omega_{\rm m}}}
\newcommand{\Ob}{\ensuremath{\Omega_{\rm b}}}

\newcommand{\ns}{\ensuremath{n_{\rm s}}}
\newcommand{\As}{\ensuremath{A_{\rm s}}}

\newcommand{\Vpeak}{\ensuremath{V_{\rm{peak}}}}
\newcommand{\Sinhagad}{\texttt{Sinhagad}}
\newcommand{\Sahyadri}{\texttt{Sahyadri}}

\usepackage[normalem]{ulem}

\title{\Sahyadri: A simulation suite for the cosmology dependence of the Cosmic Web}

\author[a,1]{Saee Dhawalikar,\note{Corresponding author.}}
\author[b]{Shadab Alam,}
\author[a]{Aseem Paranjape,}
\author[c]{and Arka Banerjee}
\affiliation[a]{Inter-University Centre for Astronomy \& Astrophysics,\\ Ganeshkhind, Post Bag 4, Pune 411007, India}
\affiliation[b]{Tata Institute of Fundamental Research, Homi Bhabha Road, Mumbai 400005, India }
\affiliation[c]{Department of Physics, Indian Institute of Science Education and Research,\\ Homi Bhabha Road, Pashan, Pune 411008, India}

\emailAdd{saee.dhawalikar@iucaa.in}
\emailAdd{shadab.alam@tifr.res.in}
\emailAdd{aseem@iucaa.in}
\emailAdd{arka@iiserpune.ac.in}

\abstract{We present \Sahyadri, a suite of cosmological $N$-body simulations designed to enable precision studies of the low-redshift Universe with next-generation spectroscopic surveys.
\Sahyadri\ includes systematic variations of four cosmological parameters around Planck 2018 constraints, with seed-matched initial conditions enabling cosmological parameter derivatives. It is planned to ultimately extend to six parameters.
Each simulation evolves $2048^3$ particles in a periodic box of side length $200\,h^{-1}\,{\rm Mpc}$, yielding a particle mass of $m_{\rm p} = 8.1 \times 10^{7}\,h^{-1}\,M_{\odot}$ in the fiducial Planck 2018 cosmology. This resolution enables robust identification of dark matter halos down to $M_{\rm min} = 3.2 \times 10^{9}\,h^{-1}\,M_{\odot}$, which represents a factor of $\sim$25 improvement over the AbacusSummit suite, and is over two orders of magnitude better than the Quijote and Aemulus suites. We estimate that approximately 40\% of DESI BGS galaxies at redshift $z < 0.15$—roughly 1.6 million objects—reside in halos accessible to \Sahyadri\ but beyond the reach of existing parameter-varying simulation suites.

We demonstrate \Sahyadri's capabilities through measurements of the matter power spectrum, halo mass function and power spectrum, and beyond 2-point statistics such as the Voronoi volume function and $k^{\rm th}$ nearest neighbour statistics, showing excellent agreement with theoretical predictions and significant sensitivity to $\Omega_{\rm m}$ variations. 
We implement a custom compression scheme reducing storage requirements by a factor of $\sim$3 while maintaining sub-percent clustering accuracy. 
Key data products are publicly available. 

}


\begin{document}
\maketitle
\flushbottom

\section{Introduction}
\label{sec:intro}

Ongoing and upcoming large-scale structure (LSS) surveys such as the Dark Energy Spectroscopic Instrument \citep[DESI;][]{DESICollaboration2016a}, the 4-metre Multi-Object Spectroscopic Telescope \citep[4MOST;][]{deJong2019}, the Subaru Prime Focus Spectrograph \citep[PFS;][]{Takada2014}, \textit{Euclid} \citep{EuclidCollaboration2022}, and the Vera C. Rubin Observatory's Legacy Survey of Space and Time \citep[LSST;][]{Ivezic2019} will map the Universe with unprecedented precision. These surveys will probe scales ranging from the quasi-linear to the deeply non-linear regime across a wide range of redshifts, providing access to a wealth of cosmological information. Extracting this information requires accurate theoretical predictions, robust summary statistics, and comprehensive modeling frameworks that can account for complex non-linear physics and baryonic effects.

Cosmological simulations have become indispensable tools for meeting this challenge. Over the past two decades, numerous simulation suites have been developed with diverse objectives and specifications—spanning different volumes, mass resolutions, cosmological parameter coverage, and baryonic prescriptions—tailored to specific science goals. High-resolution $N$-body and hydrodynamical simulations such as Uchuu \citep{Ishiyama2021}, SIMBA \citep{Dave2019}, EAGLE \citep{Schaye2015}, Illustris \citep{Vogelsberger2014}, IllustrisTNG \citep{Nelson2019,Pillepich2018}, the Millennium suite \citep{Springel2005,Boylan-Kolchin2009}, Bolshoi \citep{Klypin2011}, and MultiDark \citep{Klypin2016} have revolutionized our understanding of galaxy formation and evolution. However, these simulations typically explore a fixed or limited set of cosmological parameters, restricting their utility for constraining cosmology through LSS observations. The CAMELS suite \cite{camels-2021} bridges these categories by varying both cosmological and astrophysical parameters at a mass resolution comparable to that considered in the present work, but with volumes limited to $(50\,h^{-1}\,{\rm Mpc})^3$ which is 64 times smaller than what we use below, thus restricting large-scale structure and sample variance studies.

Conversely, simulation suites designed explicitly for cosmological inference—including Quijote \citep{Villaescusa-Navarro2020}, Aemulus \citep{DeRose2019,McClintock2019}, Dark Quest \citep{Nishimichi2019}, Abacus Cosmos \citep{Garrison2018}, and AbacusSummit \citep{Maksimova2021}—systematically vary cosmological parameters to enable derivative calculations and emulator construction. Table~\ref{tab:suite_comparison} summarizes the specifications of these suites. 
While these efforts have substantially advanced cosmological modeling, they are constrained by practical trade-offs between volume, resolution, and parameter coverage. For instance, with a particle mass of $m_{\rm p} = 2.0 \times 10^{9}\,h^{-1}\,M_{\odot}$, AbacusSummit—currently the highest-resolution suite with comprehensive parameter coverage—resolves halos down to $M_{\rm min} \approx 8.0 \times 10^{10}\,h^{-1}\,M_{\odot}$ using a conservative 40-particle threshold. Similarly, Quijote ($m_{\rm p} = 6.6 \times 10^{11}\,h^{-1}\,M_{\odot}$), Aemulus ($m_{\rm p} = 3.5 \times 10^{10}\,h^{-1}\,M_{\odot}$), Dark Quest ($m_{\rm p} = 1.0 \times 10^{10}\,h^{-1}\,M_{\odot}$), and Abacus Cosmos ($m_{\rm p} = 4.0 \times 10^{10}\,h^{-1}\,M_{\odot}$ and $1.0 \times 10^{10}\,h^{-1}\,M_{\odot}$) resolve halos down to approximately $M_{\rm min} \approx 2.6 \times 10^{13}$, $1.4 \times 10^{12}$, $4.0 \times 10^{11}$, and $1.6 \times 10^{12}$/$4.0 \times 10^{11}\,h^{-1}\,M_{\odot}$, respectively. In addition, most AbacusSummit simulations terminate at $z = 0.1$, whereas \Sahyadri\ continues to $z = 0$, capturing the full low-redshift range relevant for galaxy surveys such as DESI BGS. As we show below, these mass thresholds remain insufficient for comprehensively modeling the faint galaxy populations targeted by state-of-the-art low-redshift spectroscopic surveys \cite{Alam_2021}. 

\begin{table*}
\centering
\begin{tabular}{lccccc}
\hline\hline
Suite & $L_{\rm box}$ & $N_{\rm part}$ & $m_{\rm p}$ & $M_{\rm min}$ (40 particles) \\
 & $[h^{-1}\,{\rm Mpc}]$ &  & $[h^{-1}\,M_{\odot}]$ & $[h^{-1}\,M_{\odot}]$  \\
\hline
Quijote & 1000 & $512^3$ & $6.6 \times 10^{11}$ & $2.6 \times 10^{13}$  \\
Aemulus & 1050 & $1400^3$ & $3.5 \times 10^{10}$ & $1.4 \times 10^{12}$  \\
Dark Quest & 1000 & $2048^3$ & $1.0 \times 10^{10}$ & $4.0 \times 10^{11}$  \\
Abacus Cosmos & 1100/720 & varies & $4.0 \times 10^{10}/1.0\times10^{10}$ & $1.6 \times 10^{12}/4.0\times10^{11}$  \\
AbacusSummit & 2000 & $6912^3$ & $2.0 \times 10^{9}$ & $8.0 \times 10^{10}$ \\
\hline
\textbf{Sahyadri} & \textbf{200} & $\mathbf{2048^3}$ & $\mathbf{8.1 \times 10^{7}}$ & $\mathbf{3.2 \times 10^{9}}$ \\
\hline
\end{tabular}
\caption{Comparison of cosmological simulation suites with parameter variations. $M_{\rm min}$ denotes the minimum halo mass resolved with at least 40 particles. \Sahyadri\ achieves a factor of $\sim$25 improvement in mass resolution compared to AbacusSummit, the previous highest-resolution suite with comprehensive cosmological parameter coverage.} 
\label{tab:suite_comparison}
\end{table*}

In this paper, we present \Sahyadri, an $N$-body simulation suite specifically designed to address the resolution requirements of current and next-generation low-redshift spectroscopic surveys while maintaining the capability to compute cosmological derivatives.  \Sahyadri\ builds upon the \Sinhagad\ pilot suite introduced in \cite{dhawalikar2025stabilizingsimulationbasedcosmologicalfisher}, increasing the particle count by a factor of 512 (from $256^3$ to $2048^3$ in the same $200\,h^{-1}\,{\rm Mpc}$ volume) to achieve the mass resolution necessary for comprehensive modeling of low-redshift galaxy samples. With a particle mass of $m_{\rm p} = 8.1 \times 10^{7}\,h^{-1}\,M_{\odot}$ in its default cosmology, \Sahyadri\ resolves dark matter halos down to $M_{\rm min} = 3.2 \times 10^{9}\,h^{-1}\,M_{\odot}$ (40 particles), representing a factor of $\sim$25 improvement in mass resolution compared to the default AbacusSummit simulation, and over 2 orders of magnitude better than Quijote and Aemulus. This enhanced resolution is particularly critical for modeling the DESI Bright Galaxy Survey \citep[BGS;][]{Hahn2023}, which targets galaxies at $z \lesssim 0.5$ with peak number density at $z < 0.2$, and similar low-redshift samples from 4MOST \citep{Yildiz2020}. Table~\ref{tab:suite_comparison} compares \Sahyadri\ with the existing cosmological variation simulation suites.

Using empirical stellar-to-halo mass relations from abundance matching \citep{Behroozi2019} and the stellar mass-to-luminosity relation of \cite{Bell2003}, we find that approximately 40\% of BGS galaxies at $z < 0.15$ reside in halos below the AbacusSummit resolution limit but within \Sahyadri's reach. This corresponds to roughly 1.6 million galaxies out of a total BGS sample of $\sim$4 million in this redshift range that can be modeled in \Sahyadri\ but are inaccessible to AbacusSummit (see Alam et al., in preparation, for detailed methodology).

\begin{figure}
    \centering
    \includegraphics[width=\textwidth,]{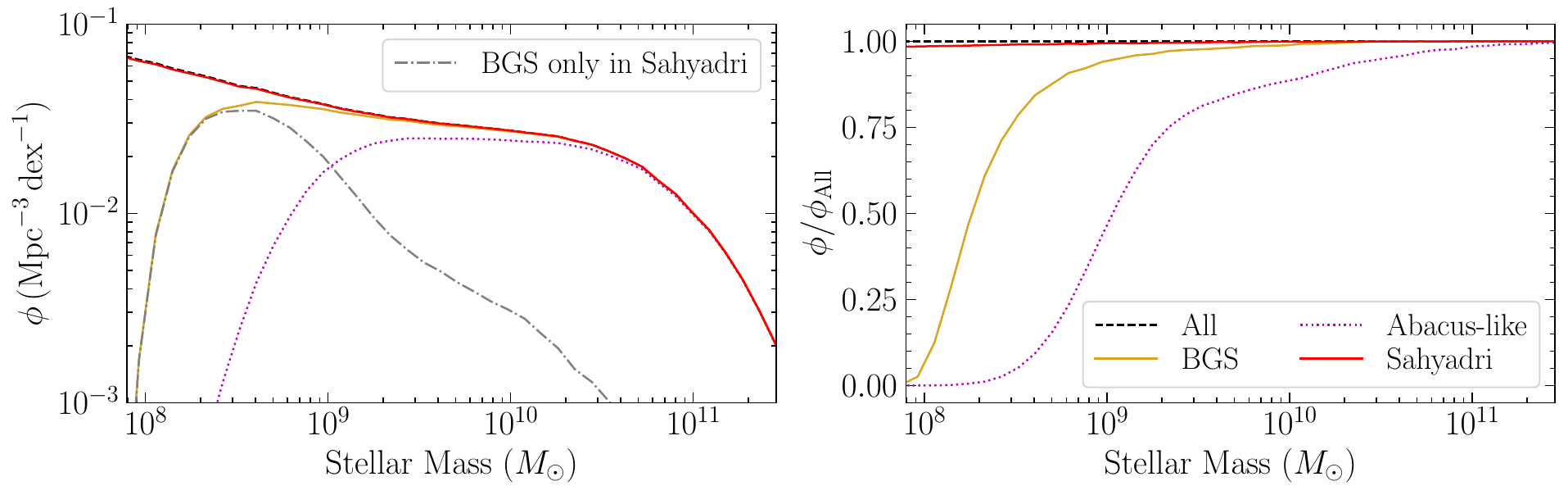}
    \caption{Stellar mass function using a stellar-to-halo mass relation and a conditional $r$-band magnitude distribution at redshift $z=0.15$ based on Alam et al. (in prep). We show that the default \Sahyadri\ simulation allows us to  model much lower stellar masses as compared to AbacusSummit which misses $\sim 40\%$ of galaxies in BGS at these low redshifts.}
    \label{fig:smf}
\end{figure}

Figure~\ref{fig:smf} quantifies this improvement by comparing the stellar mass functions from the Galaxy And Mass Assembly \citep[GAMA;][]{Driver2011} survey with the resolution limits of AbacusSummit and \Sahyadri. The left panel displays the stellar mass function, while the right panel shows the completeness fraction relative to the full stellar mass function (black dashed curve). Results from \Sahyadri\ and AbacusSummit are shown in red and magenta, respectively, while the yellow curve corresponds to the DESI BGS sample. The grey dash–dotted curve in the left panel shows the stellar mass function of BGS galaxies that are resolved in \Sahyadri\ but absent in AbacusSummit. For AbacusSummit, completeness drops below 80\% at stellar mass $M_{\star} \lesssim 3 \times 10^{9}\,M_{\odot}$, whereas DESI-BGS maintains the same completeness down to $M_{\star} \lesssim 3 \times 10^{8}\,M_{\odot}$ and \Sahyadri\ is able to capture this more than an order of magnitude improvement. The BGS spectroscopic sample is within \Sahyadri's optimal resolution regime while extending significantly beyond AbacusSummit's capabilities.

This mass resolution opens up new avenues for precision studies of the low-mass galaxy population, including the faint end of the stellar mass function \citep{Baldry2012}, environmental quenching in low-mass satellites \citep{Wetzel2013}, and AGN activity in dwarf galaxies \citep{Reines2013}. Furthermore, the low-redshift Universe ($z < 0.4$) represents a regime of considerable contemporary interest, not only because surveys achieve their highest number densities there, but also because several outstanding cosmological tensions may have roots in low-redshift physics. Both the Hubble tension, the discrepancy between local distance ladder measurements \citep{Riess2022} and CMB-inferred values \citep{Planck18-VI-cosmoparam}, as well as hints of evolving dark energy from type Ia supernovae and BAO measurements \citep{DESIDR1_cosmo, DESIDR2_cosmo} suggest potential new physics operating at $z \lesssim 0.4$. Current analyses of large-scale structure at these redshifts have approached cosmic variance limits in two-point statistics given available volumes \citep{Wadekar2020,DESIDR1_cosmo}. Extracting additional cosmological information necessitates probing non-linear scales and higher-order statistics, both of which require the combination of high mass resolution and cosmological parameter variations that \Sahyadri\ provides.

The simulation volume of $(200\,h^{-1}\,{\rm Mpc})^3$ represents a carefully chosen compromise. While smaller in volume than suites optimized for Baryon Acoustic Oscillation (BAO) studies, this volume is sufficient for convergence in large-scale clustering statistics such as the linear halo bias $b_1$ (see Section~\ref{subsec:halos_vahc} and Appendix~\ref{app:b1-voldep}), while enabling the enhanced mass resolution critical for low-redshift galaxy surveys.

The remainder of this paper is organized as follows. Section~\ref{sec:specs} describes the simulation suite specifications, including cosmological parameters and their variations. Section~\ref{subsec:simulations} details the numerical methods, particle data compression scheme, and computational requirements. Section~\ref{subsec:halos_vahc} presents the halo catalogs, cleaning procedures, and sample definitions used throughout our analyses. Section~\ref{sec:highlights} showcases the scientific capabilities of \Sahyadri\ through visualizations of the cosmic web and studies of the cosmological dependence of the matter and halo power spectrua, halo mass function, Voronoi volume function (VVF), $k^{\rm th}$ nearest neighbour ($k$NN) statistics, and correlations between halo environment and internal properties at $z=0$ and $z=1$. We conclude in Section~\ref{sec:summary} with a summary of our results and an outlook for future applications. Appendix~\ref{App:fitfuncs} provides detailed comparisons of our clustering measurements with theoretical predictions. Appendix~\ref{app:vvf-scale} collects some useful results regarding VVF statistics, while Appendix~\ref{app:b1-voldep} presents convergence tests quantifying the effects of finite simulation volume on large-scale bias measurements.

\section{Suite Specifications}
\label{sec:specs}

\begin{table}[t]
    \centering
    \begin{tabular}{lccc}
    \hline\hline
        Parameter & Fiducial value ($\theta_{\rm{f}}$) & Variation magnitude ($\Delta$) & Status \\
        \hline
        $\Omega_{\rm m}$ &  $0.3138$ & $0.05\,\Omega_{\rm{m,f}}$ & Complete \\
        $n_{\rm s}$ &  $0.9649$ & $0.05\,n_{\rm{s,f}}$ & Complete \\
        $h$ &  $0.6736$ & $0.05\,h_{\rm{f}}$ & Complete\\
        $A_{\rm s}$ &  $2.0989\times 10^{-9}$ & $0.1\,A_{\rm{s,f}}$ & Complete \\
        $w_{\parallel }$ &  $0$ & $0.125$ & In progress\\
        $\Omega_{\rm k}$ &  $0.0$ & $0.05$ & Planned\\
        \hline
    \end{tabular}
    \caption{Fiducial cosmological parameters and variation magnitudes for the \Sahyadri\ simulation suite. The subscript ``f'' denotes fiducial values. Each parameter is varied independently while maintaining all other parameters at their fiducial values. }
    \label{tab:cos_params}
\end{table}

The \Sahyadri\ simulation suite adopts fiducial cosmological parameters consistent with the Planck 2018 analysis \citep{Planck18-VI-cosmoparam}. Table~\ref{tab:cos_params} lists the fiducial parameter values ($\theta_{\rm f}$) and their respective variation magnitudes ($\Delta$). This configuration enables Fisher matrix analyses, wherein each of the six cosmological parameters is varied individually while holding the remaining parameters fixed at their fiducial values. 

\begin{figure}
    \centering
    \includegraphics[width=\linewidth]{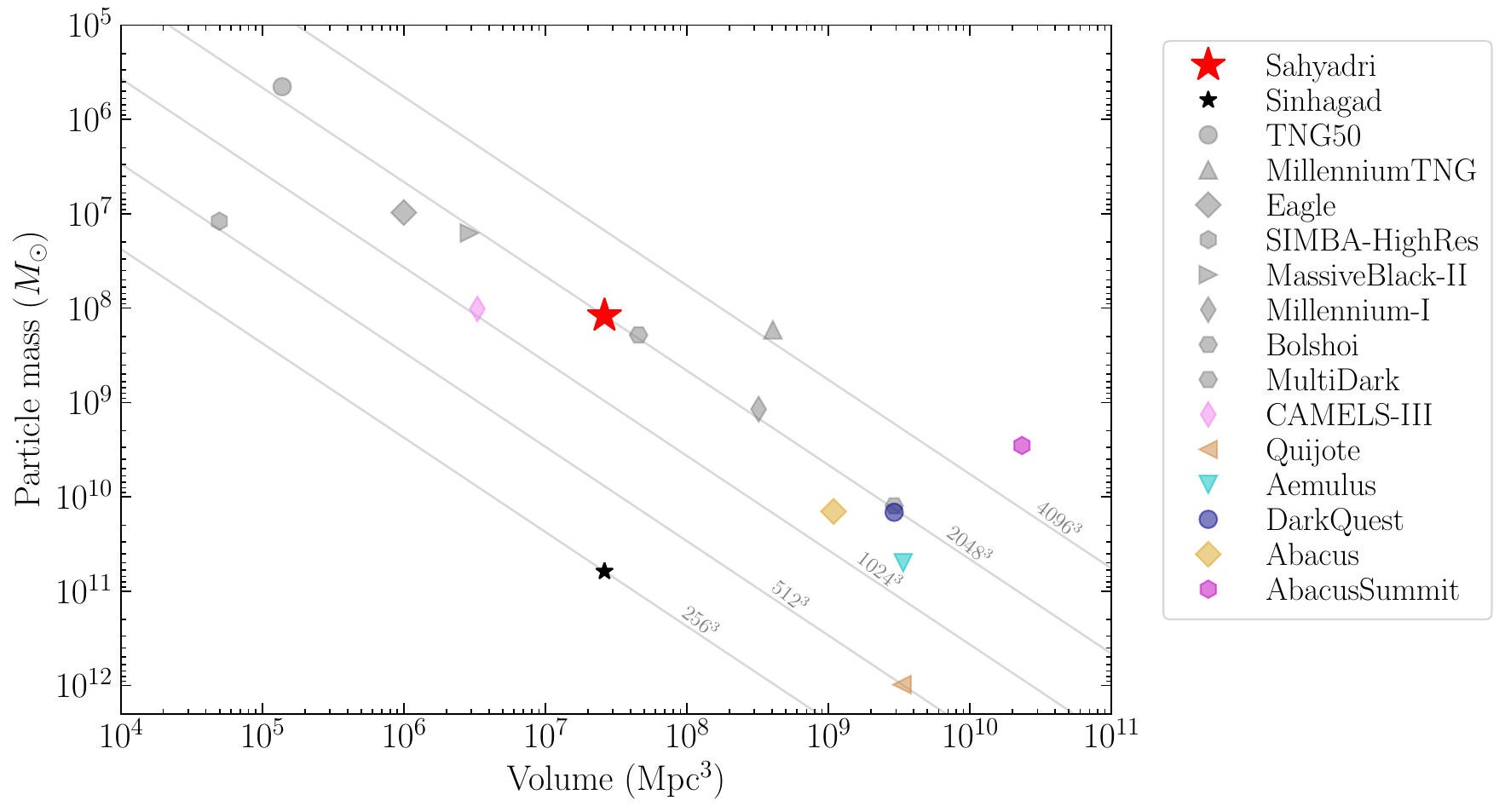}
    \caption{Comparison of the \Sahyadri\ simulation suite with other large-scale structure simulation efforts. Grey lines indicate simulations with constant particle number assuming Planck 2018 cosmology. Coloured points highlight suites with cosmology variations. The red (black) star marks the \Sahyadri\ (\Sinhagad) simulation configuration, highlighting the state-of-the-art mass resolution achieved by our suite. }
    \label{fig:intro}
\end{figure}

Figure~\ref{fig:intro} places the \Sahyadri\ suite in context with other contemporary simulation efforts, illustrating the trade-off between box size and mass resolution. The grey lines represent simulations with constant particle numbers assuming Planck 2018 cosmology, while the black (red) star denotes the \Sahyadri\ (\Sinhagad) configuration.

The matter density parameter \Om\ is precisely constrained in $\Lambda$CDM, with Planck 2018 yielding $\Omega_{\rm m} = 0.315 \pm 0.007$ \citep{Planck18-VI-cosmoparam}. However, recent analyses reveal tensions in $\Omega_{\rm m}$ across different probes.  Interesting discrepancies have emerged between weak lensing (CMB and galaxy) and primary CMB measurements along the \Om\ axis (see Figure 12 and associated discussion in \citep{Hang_2020}), while \citep{Ishiyama2024} demonstrated that the BAO feature locations are nearly identical between Planck 2018 $\Lambda$CDM and Planck 2018 with DESI-preferred values of the dark energy equation-of-state parameters $(w_0, w_a)$. This indicates that apparent dark energy evolution cannot be attributed to dynamics alone but reflects differences in \Om\ and the sound horizon $r_{\rm d}$. These findings highlight the critical role of precise non-linear structure modeling at $z \lesssim 1$, where \Om\ variations produce interesting signatures.

Therefore, the analyses in this work focus on \Om\ variations, demonstrating a methodology that is applicable to the complete suite. At the time of this publication, we have completed the fiducial cosmology and full variations for \Om, $h$, \ns\, and \As, with the remaining parameters ($w_\parallel$ and $\Omega_k$) in progress. In total, the \Sahyadri\ suite will comprise 13 simulations: one fiducial and two variations for each of the six cosmological parameters. For the dark energy equation of state, we define $w_\parallel$ along the CMB degeneracy $w_a = A(w_0 + 1)$ (where $A = -3.68$) via $w_0 = -1 + w_\parallel/\sqrt{1 + A^2}$ and $w_a = A w_\parallel/\sqrt{1 + A^2}$ (setting $w_\perp = 0$), allowing efficient sampling of $D_M(z_\star)$-consistent models (see Lodha et. al. in prep. for more details).

\subsection{Simulations}
\label{subsec:simulations}

All simulations in the \Sahyadri\ suite were executed in a periodic comoving box of side length $L_{\rm box} = 200\,h^{-1}\,{\rm Mpc}$ containing $2048^3$ dark matter particles. This corresponds to a particle mass of $m_{\rm p} = 8.1 \times 10^{7}\,h^{-1}\,M_{\odot}$ in the fiducial cosmology. At this resolution, as mentioned earlier, halos with masses $M > 3.2\times10^{9}\,h^{-1}\,M_{\odot}$ are resolved with more than 40 particles, satisfying standard convergence criteria for halo identification and statistical analysis. We note that massive neutrinos are not included in these simulations.

\subsubsection{Numerical Methods}

We employed the massively parallel tree-PM code \textsc{gadget-4} \citep{Gadget4}\footnote{\url{https://wwwmpa.mpa-garching.mpg.de/gadget4/}} with the ``lean'' memory configuration. The simulations adopted a comoving Plummer-equivalent force softening length of $\epsilon = L_{\rm box}/(30 \times 2048) \approx 3.26\,h^{-1}\,{\rm kpc}$, corresponding to $1/30$ of the mean inter-particle spacing. Long-range gravitational forces were computed using a particle-mesh (PM) algorithm on a $4096^3$ grid, ensuring Nyquist sampling of the particle distribution.

Initial conditions were generated at $z = 49$ using second-order Lagrangian perturbation theory \citep[2LPT;][]{Scoccimarro1998} via the \textsc{N-GenIC} code integrated into \textsc{gadget-4}. The linear matter power spectrum was computed using the Boltzmann code \textsc{class} \citep{Blas2011,class-II}.\footnote{\url{https://github.com/lesgourg/class\_public}} To enable controlled cosmological derivatives, all parameter-varied simulations employed identical random number seeds (i.e., matching initial phases) as the fiducial run, ensuring that differences arise solely from the cosmological parameter changes. Although this is not sufficient to control the gravitationally induced stochasticity due to phase-mixing at small scales, which affects statistical probes involving halos \citep[e.g.,][]{coulton+23}, such effects can be addressed using additional post-processing optimization strategies as described in \cite{dhawalikar2025stabilizingsimulationbasedcosmologicalfisher}. This allows \Sahyadri\ to be used reliably for estimating derivatives of small-scale observables with respect to cosmological parameters.

Each simulation generated 101 snapshots between $z=12$ and $z=0$, uniformly spaced in scale factor $a = (1+z)^{-1}$ with $\Delta a = 0.01$. While the full snapshot sequence is available, the analyses presented in this paper focus primarily on $z=0$ and $z=1$. Each simulation was executed on 12 compute nodes with 32 cores each, utilizing approximately 250 GB of memory per node and requiring $\sim 0.43$ million CPU hours per run.

\subsubsection{Particle Data Compression}

Storing full particle information for all snapshots and cosmological variations presents significant data management challenges. A single uncompressed snapshot contains $\sim$150 GB of position, velocity, and particle ID data, yielding $\sim$15 TB per simulation for 101 snapshots. Inspired by the compression strategies employed in the AbacusSummit suite \citep{AbacusSummit}, we implemented a custom compression scheme that reduces storage requirements by a factor of $\sim$3 while maintaining sufficient precision for most scientific applications.

Our compression algorithm proceeds as follows:
\begin{enumerate}
    \item We tessellate the simulation box with a uniform $400^3$ grid, assigning each particle to the corresponding cell.
    \item Particles are sorted by their grid cell index, improving cache efficiency for subsequent I/O operations.
    \item Within each cell, particle positions are stored relative to the cell center using 8-bit integers per coordinate, providing a spatial resolution of $\Delta x = L_{\rm box}/(400 \times 256) \approx 1.95\,h^{-1}\,{\rm kpc}$, well below the force softening scale.
    \item The particle count per cell is stored as a 12-bit unsigned integer, accommodating up to 4095 particles per cell. For the rare high-density cells exceeding this limit, overflow particles are stored separately with full precision.
    \item Particle velocities are mapped to the range $v \in [-5000, 5000]\,{\rm km\,s^{-1}}$ and quantized to 12-bit precision, yielding a velocity resolution of $\Delta v \approx 2.44\,{\rm km\,s^{-1}}$, adequate for computing velocity statistics and identifying virialized structures.
    \item Particle IDs are stored without compression to enable exact particle tracking across snapshots.
\end{enumerate}

This scheme reduces position storage from 96 bits (three 32-bit floats) to 24 bits per particle, and velocities from 96 to 36 bits, while maintaining sub-percent accuracy for clustering measurements. The grid cell metadata contributes negligibly to the total file size. 

To facilitate flexible data access and future storage optimization, we partition each snapshot into ten subsamples split into 1, 3, 5, 9, 10, 12, 15, 20 and 25 percent particles, respectively, stored in separate files. This structure enables users to download partial datasets tailored to their resolution requirements and allows straightforward deletion of higher-sampling files if storage becomes limited.

The compressed snapshot format requires 56 GB per snapshot without particle IDs ($\sim$5.6 TB for 101 snapshots per simulation) or 85 GB with IDs ($\sim$8.6 TB per simulation). For the full planned \Sahyadri\ suite with variations of six cosmological parameters (each with positive and negative deviations plus the fiducial model), the total snapshot storage footprint is approximately 112 TB, making public data release feasible through modern cloud storage infrastructure.

\subsection{Halo and environmental catalogs}
\label{subsec:halos_vahc}
Dark matter halos in the simulations were identified using the six-dimensional phase-space Friends-of-Friends algorithm implemented in \textsc{rockstar} \citep{Rockstar1023},\footnote{\url{https://bitbucket.org/gfcstanford/rockstar/}} modified by us to allow for $\Omega_{\rm k}$ variations (which are not discussed in this paper). Halo merger trees were constructed with the \textsc{consistent-trees} code \citep{Consistent_trees2013}.\footnote{\url{https://bitbucket.org/pbehroozi/consistent-trees/}} The high density of stored snapshots ensures reliable tracking of merger and accretion relations between halos across cosmic time.

Alongside the primary halo catalogs produced by \textsc{rockstar} and \textsc{consistent-trees}, we calculate a number of environmental quantities for each halo. These are recorded separately in a `value added' halo catalog for each snapshot and include the following.

\begin{itemize}
\item Eigen-values $\{\lambda_1,\lambda_2,\lambda_3\}$ of the halo-centric tidal tensor $T_{ij}(\xx_h;R)$ smoothed at scales $R \in \{2,4,6,8\}\times R_{\rm 200b}$. (Here $\xx_h$ denotes the spatial location of halo $h$.)
\item Eigen-values $\{\lambda_1,\lambda_2,\lambda_3\}$ of $T_{ij}(\xx_h;R)$ smoothed at the fixed scales $R\in\{2,3,5\}$ \Mpch.
\item Eigen-values $\{h_1,h_2,h_3\}$ of the halo-centric density Hessian $H_{ij}(\xx_h;R)$ smoothed at the fixed scales $R\in\{2,3,5\}$ \Mpch.
\item Halo-by-halo bias $b_1(h)$.
\end{itemize}

The technique for calculating the halo-centric tidal tensor and density Hessian is the same as described in \cite{phs18}. Briefly, the smoothed tidal tensor $T_{ij}(\xx;R)=\partial_i\partial_j\psi(\xx;R)$ is estimated by first inverting the unsmoothed Poisson equation $\nabla^2\psi=\delta$ in Fourier space and multiplying by the Fourier transform $W(kR)$ of the relevant (Gaussian) smoothing kernel, before inverse Fourier transforming. Here, $\delta$ is the matter density contrast estimated using cloud-in-cell (CIC) smoothing on a $1024^3$ grid, with Fourier transform $\delta_{\kk}$, so that
\be
T_{ij}(\kk)=\frac{k_ik_j}{k^2}\delta_{\kk}\,\longrightarrow T_{ij}(\xx;R) = {\rm F.T.}\left[T_{ij}(\kk)W(kR)\right]\,.
\ee
For the halo-scaled smoothing radius values $R\propto R_{\rm 200b}$, we first estimate the tensor on a range of fixed smoothing scales and then interpolate to the scale of interest. For the fixed smoothing scales $\{2,3,5\}$ \Mpch, the tensor is calculated by directly smoothing at the respective scale. 
In each case, $T_{ij}(\xx;R)$ is finally interpolated to the halo locations $\xx\to\{\xx_h\}$.

Below, we will showcase \Sahyadri\ results using the \emph{tidal anisotropy} variable $\alpha$ introduced by \cite{phs18} and defined as
\begin{equation}
\alpha \equiv \sqrt{q^2}/(1+\delta)\,,
\label{eq:alpha-def}
\end{equation}
where $q^2=\frac12\left[(\lambda_1-\lambda_2)^2+(\lambda_1-\lambda_3)^2+(\lambda_2-\lambda_3)^2\right]$ is the halo-centric tidal shear and $\delta=\lambda_1+\lambda_2+\lambda_3$ is the halo-centric density contrast, with $\{\lambda_i\}$ in this case being the tidal tensor eigen-values evaluated at smoothing scale $R=4R_{\rm 200b}$.

The density Hessian $H_{ij}$ is constructed similarly to the tidal tensor in Fourier space at the three fixed smoothing scales and then inverse Fourier transformed and interpolated to the halo locations:
\be
H_{ij}(\xx;R)=\partial_i\partial_j\delta(\xx;R)={\rm F.T.}\left[-k_ik_j\delta_{\kk}W(kR)\right]\,\longrightarrow H_{ij}(\xx_h;R)\,.
\ee
The halo-by-halo bias $b_1(h)$ is calculated as a suitably weighted, sharp-$k$ filtered halo-centric density using the method described by \cite{VVF2020}, which is an improved version of the technique originally introduced by \cite{phs18}. 
Schematically, for halo $h$,
\be
b_{1}(h)=\sum_{{\rm low-}k}N_k\avg{\e{i\kk\cdot\xx_h}\delta^\ast_{\kk}}_k\,/\,\sum_{{\rm low-}k}N_k\,P_{\rm mm}(k)\,,
\ee
where $\e{i\kk\cdot\xx_h}$ denotes an inverse-CIC weighted phase factor centered on the halo location $\xx_h$, $P_{\rm mm}(k)$ is the matter auto power spectrum, the angular brackets indicate an angle-average over all \kk\ modes in a bin of $k=|\kk|$, $N_k$ is the number of these modes and the sums are over low-$k$ modes, typically restricted to $k<0.1\hMpc$. We refer the reader to \cite{phs18,VVF2020} for further details of the procedure, including tests and convergence studies. The quantity $b_1(h)$ has the property that its arithmetic mean over a sample of halos is the same as the traditional definition of linear halo bias of the sample as a ratio of the halo-matter cross power spectrum and the matter auto power spectrum. The halo and value added catalogs were compressed by storing them in FITS format, achieving a factor $\sim4$ gain relative to the native ASCII.

Finally, our default post-processing pipeline (written in Python) also allows for estimates of the matter auto, halo-matter cross and halo auto power spectra and mass functions for halos resolved with $\geq40$ particles, as well as beyond 2-point statistics such as the Voronoi volume function (VVF) and $k^{\rm th}$ nearest neighbour ($k$NN) statistics (see below). In the interest of reproduceability, our entire pipeline for performing the simulations, generating halo and environment catalogs and other post-processing products, along with the data compression described above, is publicly available at \url{https://github.com/a-paranjape/sahyadri-sandbox}. The Pythonic post-processing code, in particular, is modular and easily extendable to include other statistics.

The final compressed data products occupy approximately 9 TB of storage for each cosmology variation, about $96\%$ of which corresponds to the snapshot storage. The complete suite thus occupies approximately 117 TB of data. The average run time of each variation, including halo finding and post-processing, was approximately 0.44 million CPU hours, with the `$+$' (`$-$') variations typically being slower (faster) than the default box. The simulations and analysis were performed on the Pegasus cluster at IUCAA, Pune.\footnote{\url{http://hpc.iucaa.in}}

\subsection{Halo samples for present analysis}
\label{subsec:halo_samples}
For all analyses in this paper,  the halo catalogs are cleaned according to the degree of relaxation of each halo, quantified by $\eta \equiv 2T/|U|$, where $T$ and $U$ are the total kinetic and potential energies of the halo. Following \citep{Bett+2007}, we retain only halos with $0.5 \leq \eta \leq 1.5$, a criterion that we refer to as the QE cut. Halo mass is defined as $M_{200\rm b}$, the gravitationally bound mass enclosed within a radius $R_{200\rm b}$ corresponding to 200 times the mean matter density of the Universe.

We work with two distinct halo samples. The first, designed for calculating the VVF and $k$NN statistics, is observationally motivated. Halos are selected based on a threshold in their maximum circular velocity along the main progenitor branch of the merger tree, denoted \Vpeak\ \citep{Reddick+2013}, which correlates closely with stellar mass in subhalo abundance matching \cite{SHAM_2015, SHAM2016, SHAM_RSD2022}. Only halos resolved with at least 40 particles and satisfying the QE cut are included. The \Vpeak\  thresholds are then chosen to ensure fixed tracer number densities across simulations, with three samples at $2\times10^{-4}$, $2\times10^{-3}$, and $2\times10^{-2}\, (\rm{Mpc})^{-3}$, respectively. In the default cosmology, for $z=0$, this corresponds to \Vpeak\ thresholds of $422.5,196.6,74.6\,\kms$, respectively. The highest number density is set such that the fraction of unresolved halos (i.e., halos having $<40$ particles enclosed within $M_{\rm{200b}}$) at the corresponding \Vpeak\ threshold is $\lesssim 0.5\%$. We refer to this as the \Vpeak-selected sample below. 

The second sample is more theoretically motivated and intended for comparison with analytic fitting functions. Here, in addition to the QE cut and the 40 particle mass cut, we exclude all subhalos and retain only parent halos in order to remove 
the effects of substructure. We refer to this as the mass-selected sample below.

\section{Cosmic web studies with \Sahyadri: Highlights}
\label{sec:highlights}
As mentioned in the Introduction, \Sahyadri\ opens up multiple avenues for studies of the cosmology dependence of the small-scale distribution of faint objects. In this section, we showcase some of these possibilities by presenting visualizations of various aspects of the cosmic web defined by our observationally motivated halo samples and explicitly displaying the \Om-dependence of beyond 2-point observables such as the VVF and $k$NN statistics.

\subsection{Overview and visualizations}
\label{subsec:visualizations}

We begin by showing the redshift evolution of the dark matter density field for the fiducial cosmology, before moving on to comparisons with existing simulations and tracer-dependent visualizations of the cosmic web.
These Figures are intended to provide qualitative insight and to motivate the statistical analyses presented later in this section.

\begin{figure}
\centering
\includegraphics[width=\linewidth]{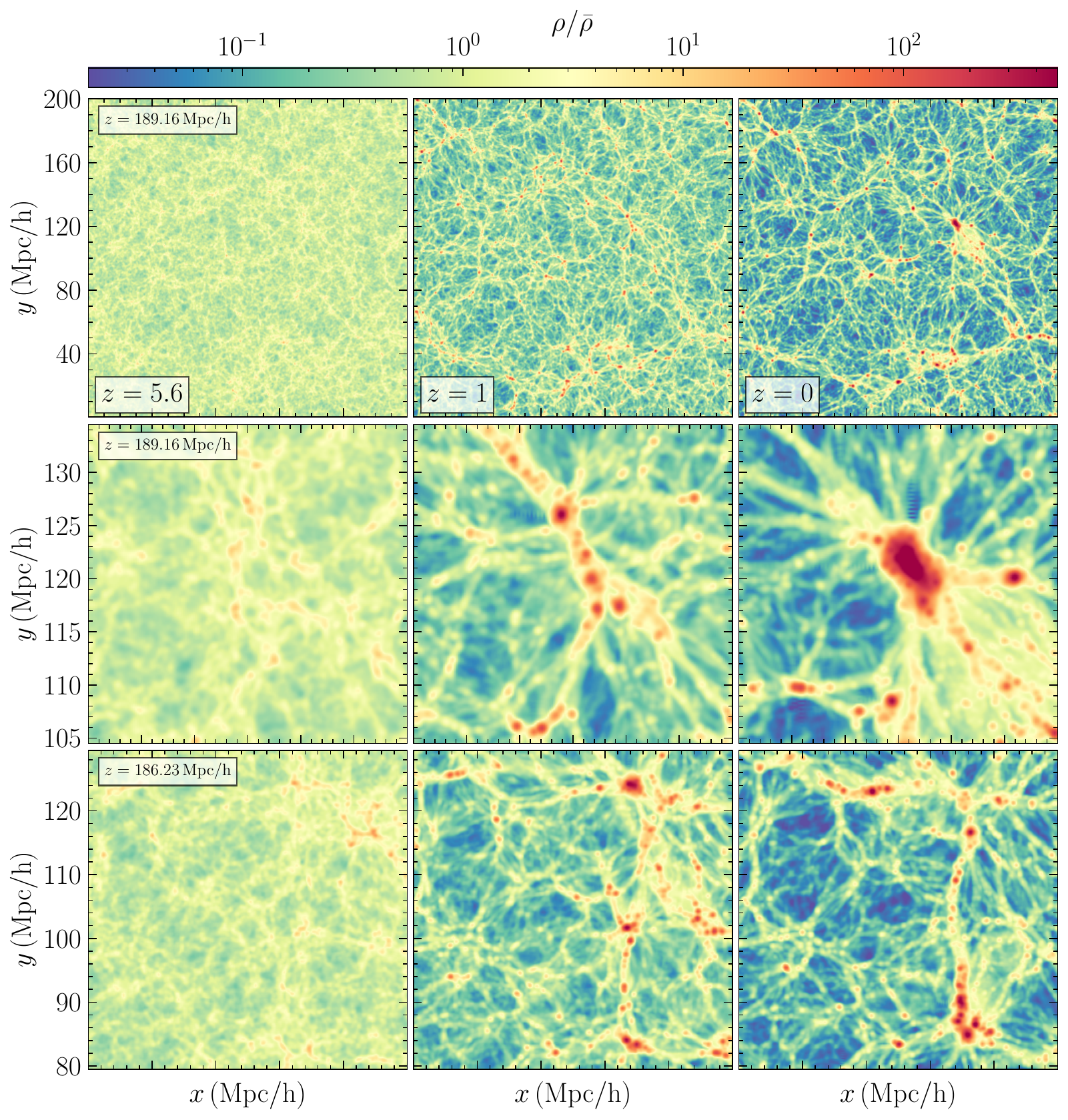}
\caption{Visualization of the evolution of the dark matter density field as a function of redshift for the fiducial cosmology. The density field is evaluated on a $1024^3$ grid and smoothed with a Gaussian kernel of radius $185\,\kpch$. The \emph{left, middle and right columns} correspond to the same single-cell spatial slices at redshift $z=5.6, 1, 0$ respectively. The \emph{top panel} shows a slice through the entire simulation box containing the most massive halo at $z=0$. The \emph{middle panel} provides a zoom-in on this halo, while the \emph{bottom panel} shows the evolution of a region that later develops into a prominent filament. The centers of the slices are indicated in each column. Each panel’s $x$- and $y$-axes span the same length, ensuring uniform tick mark size in the $x$ and $y$ directions. }
\label{fig:z_comparison}
\end{figure}

Figure \ref{fig:z_comparison} shows the evolution of the dark matter density field from high redshift to the present epoch for the fiducial cosmology. The \emph{top row} displays a slice through the full simulation volume containing the most massive halo at 
$z=0$, while the \emph{middle} and \emph{bottom rows} focus on regions that evolve into the most massive cluster in the box, and a prominent filament respectively. The dark matter density fields are estimated using CIC interpolation onto a $1024^3$ grid. The resulting field is further smoothed with a Gaussian kernel of radius $0.95$ times the grid size, corresponding to $\approx 185\,\kpch$.  Throughout, we display single slices of this density grid. The Figure highlights the well-known emergence of the cosmic web from initially diffuse density fluctuations, the growth of filaments and nodes, and the increasing density contrast due to gravitational evolution towards low redshifts.

\begin{figure}
    \centering
    \includegraphics[width=\linewidth]{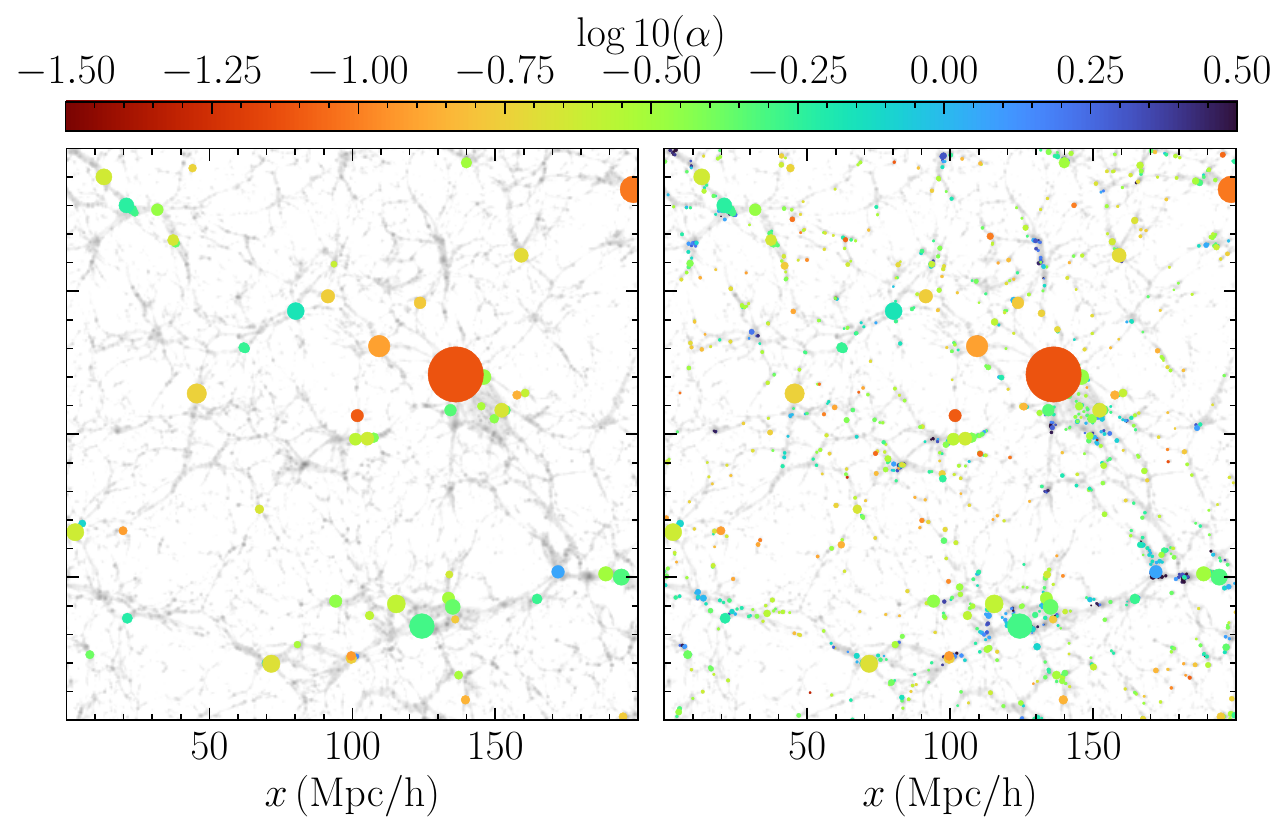}
    \caption{Comparison between AbacusSummit-like and \Sahyadri\, halos. \textit{Right panel} shows the highest number density halo sample used in this paper, with $n=2\times 10^{-2} \rm{Mpc}^{-3}$ (see section~\ref{subsec:halo_samples} for the selection criterion), from the default \Sahyadri\, simulation at $z=0$. \textit{Left panel} shows the sample of halos from the same simulation, but after degrading the particle mass to that of AbacusSummit and applying the same selection criterion, leading to $n=6.5\times 10^{-4} \rm{Mpc}^{-3}$.
    Both panels display the same spatial slice as the top right panel of Figure~\ref{fig:z_comparison}, including halos in this slice and the two neighboring slices, overplotted on the underlying density field. Circles mark halo positions, with radii equal to $4R_{\rm{200b}}$, and are colored by $\log_{10}(\alpha)$, where $\alpha$ is the tidal anisotropy defined in \eqn{eq:alpha-def}. The  $x$- and $y$-axes span the same length, ensuring uniform tick mark size in each direction. The higher resolution of \Sahyadri\ allows significantly more low-mass halos to be resolved, improving the sampling of all cosmic environments.}
    \label{fig:alpha_comparison}
\end{figure}

\begin{figure}
    \centering
    \includegraphics[width=\linewidth]{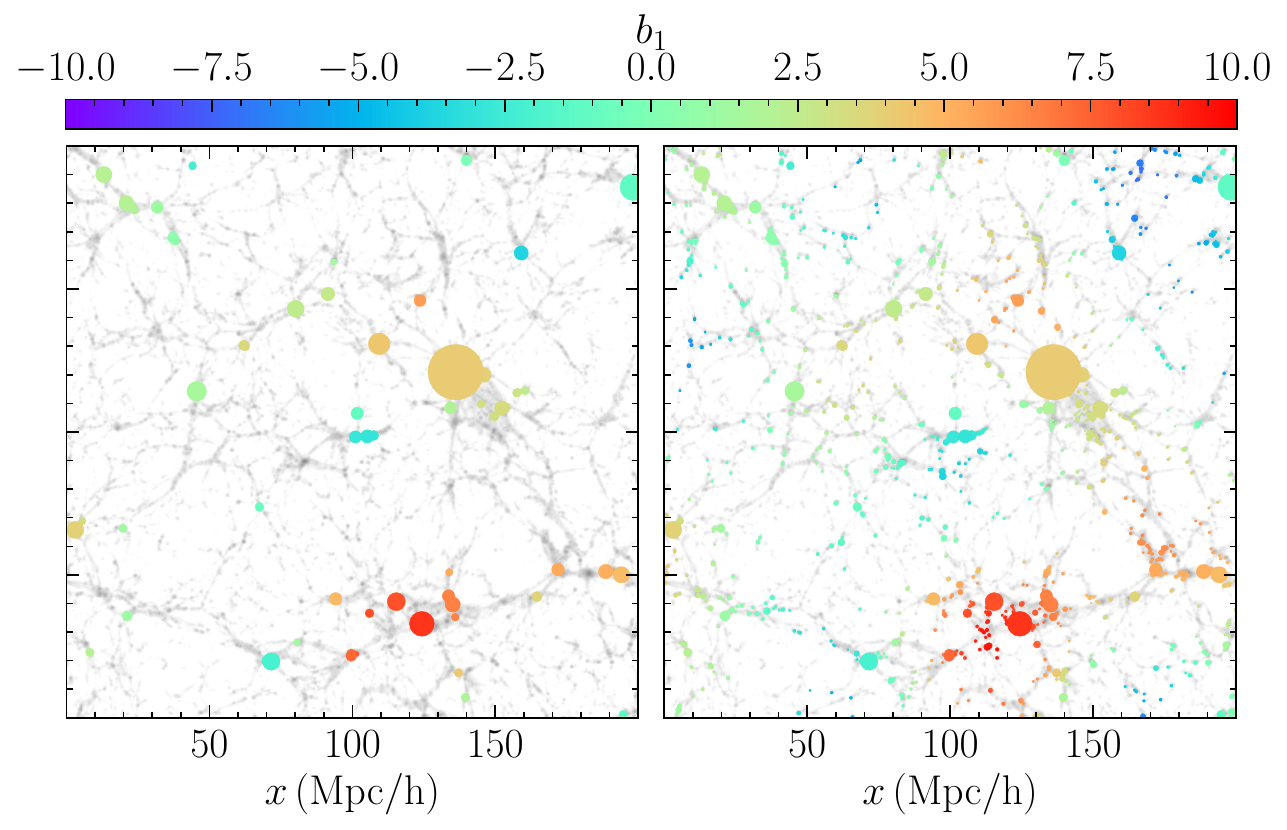}
    \caption{Same as  Figure~\ref{fig:alpha_comparison}, but with halos are coloured by the halo-by-halo bias $b_1$. We see that the most massive halo does not necessarily correspond to the largest bias, which instead traces the larger-scale density field, as established in the literature. See text for a discussion.} 
    \label{fig:b1_comparison}
\end{figure}

To visualize the impact of mass resolution on resolved halo populations, Figures \ref{fig:alpha_comparison} and \ref{fig:b1_comparison} compare a halo sample drawn from the same simulation at $z=0$, analyzed at the native \Sahyadri\ resolution, with an AbacusSummit-like sample.
The \emph{right panels} show the highest number-density \Vpeak-selected halo sample used in this work, while the respective \emph{left panels} show the corresponding sample obtained after degrading the particle mass to that of AbacusSummit and applying the same selection criterion. The increase in the abundance of low-mass halos in \Sahyadri\ is clearly visible across all environments of the cosmic web. In particular, the additional halos populate filaments, cluster outskirts, and underdense regions that are sparsely sampled in lower-resolution simulations.

Following the picture established in \citep{phs18}, Figure~\ref{fig:alpha_comparison} shows that filamentary environments are predominantly populated by halos with large tidal anisotropy $\alpha$ (equation~\ref{eq:alpha-def}).  On the other hand, large halos have very low $\alpha$, and smaller halos in their outskirts, which experience strong tidal influence, have correspondingly larger $\alpha$. Figure~\ref{fig:b1_comparison} highlights the strong environmental dependence of large-scale bias, consistent with the established view that the halo bias is not simply a monotonic function of halo mass, but is better thought of as a specific smoothing of the large-scale density environment of the halo, as discussed in \citep{phs18} and references therein.

All of these qualitative trends have been reported in earlier studies. The new aspect here is that they can be followed to substantially lower halo masses. The increased mass resolution of \Sahyadri\ results in a substantially larger population of low-mass halos across the entire web, which is directly visible in the Figures. In particular, the dense cluster region around $(x,y)=(120,30) \Mpch$ is populated by numerous high-bias halos, while extended underdense regions such as $(170,170) \Mpch$ and $(10,130) \Mpch$ exhibit a large number of strongly negative-bias halos that are largely absent at lower resolution. This leads to a smoother and more continuous representation of the cosmic web and extends the validity of these trends well beyond the mass scales previously accessible.

\begin{figure}
\centering
\includegraphics[width=\linewidth]{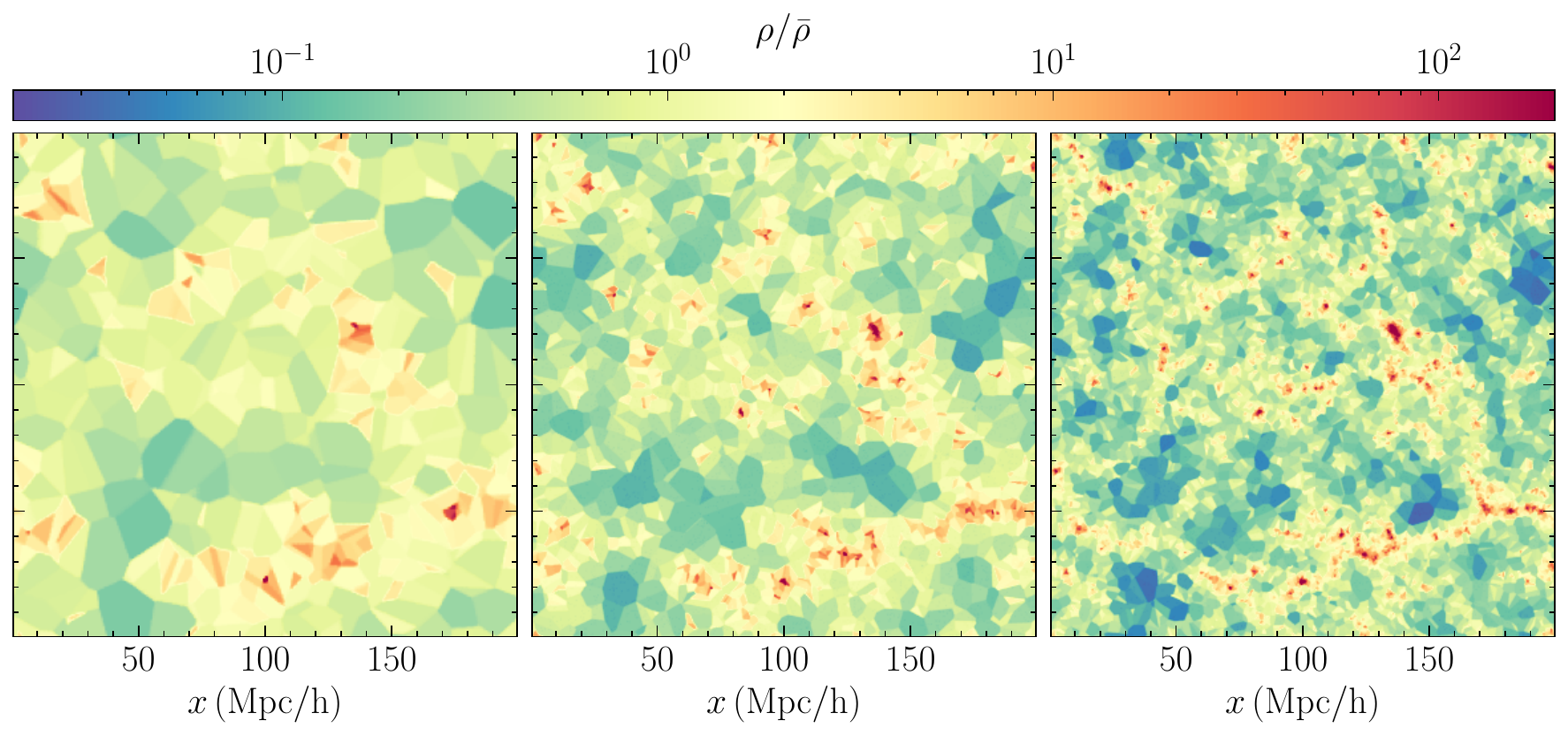}
\caption{Visualizations of single-cell slices of the tessellated density field at $z=0$. The panels from \emph{left} to \emph{right} show density estimates obtained from a Voronoi tessellation of tracers with number density $2\times 10^{-4}, 2 \times 10^{-3}, 2\times 10^{-2}\, (\rm{Mpc})^{-3}$ respectively, evaluated on a $256^3$ grid for the lowest number density and $1024^3$ grid for the two higher number densities. All the slices are centered at the same position in the simulation box, the same as the top right panel in Figure~\ref{fig:z_comparison}, and the $y$-axis spans the full box length from 0 to 200 Mpc/h. See text for details.  
}
\label{fig:voronoi_field}
\end{figure}

The role of tracer density in resolving the cosmic web is illustrated in Figure \ref{fig:voronoi_field}, which shows Voronoi-based density fields constructed from the \Vpeak-selected tracer populations with increasing number density that were described in section~\ref{subsec:halo_samples}. For the Voronoi density estimation, we adopt the Monte Carlo algorithm described in \citep{VVF2020} to compute cell densities. These densities are then mapped onto a grid using nearest-grid-point (NGP) interpolation.
As the tracer density increases, the structure of the cosmic web become progressively better resolved (cf., the top right panel of Figure~\ref{fig:z_comparison}).

Overall, these visualizations demonstrate a key capability of the \Sahyadri\ suite: the resolution of low-mass halos across a wide range of cosmic environments. This enables cosmic web analyses based on dense,  well-sampled tracer populations. In the following section, we quantify the cosmological information content of a set of clustering and environment-sensitive statistics.

\subsection{Cosmological dependence of statistics}
\label{subsec:cosmo_dependence}
We now turn to a quantitative analysis of the cosmological dependence of several dark-matter and halo-based statistics. These include both conventional measures of clustering and abundance as well as statistics that probe the geometric and environmental properties of the cosmic web.

We use the mass-selected sample for analysing the matter auto power spectrum and halo mass function, and the \Vpeak-selected samples for the halo auto power spectrum, VVF and $k$NN statistics (see section~\ref{subsec:halo_samples} for the sample definitions).
Results are presented at $z=0$ and $z=1$.

\begin{figure}
    \centering
    \includegraphics[width=\linewidth]{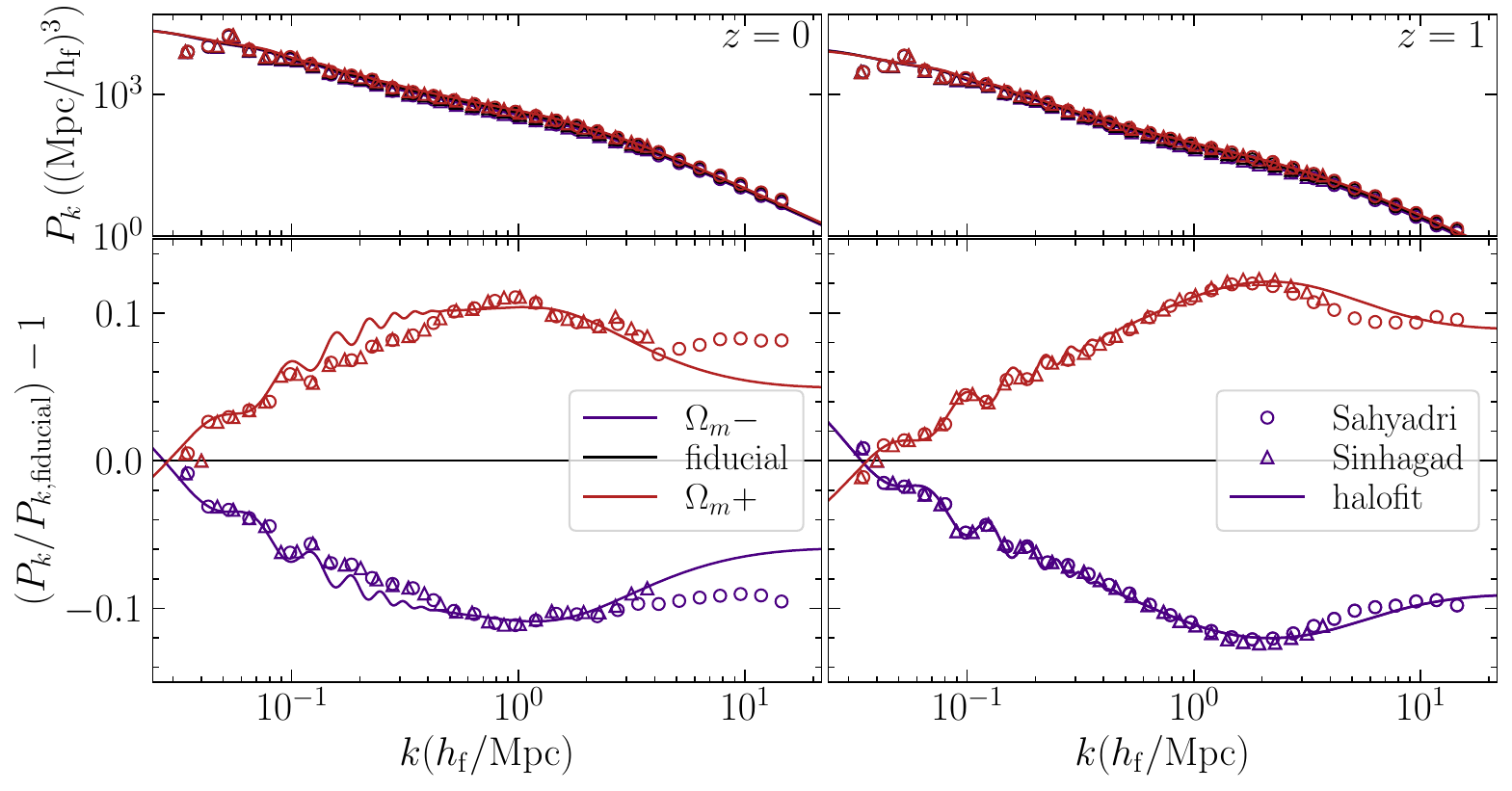}
    \caption{Variation of the matter power spectrum as a function of \Om. Bottom panels show the ratio of the matter power spectrum for each cosmology to that of the fiducial cosmology, and top panels show the full matter power spectrum. Red (blue) corresponds to the $\Om+$ ($\Om-$) variations, while black represents the fiducial cosmology. The left (right) panel shows the results at $z=0$ ($z=1$). Solid lines indicate the Halofit expectations, and circles (triangles) denote results from \Sahyadri\ (\Sinhagad). The simulations show good agreement with the model, particularly at $z=1$.}
    \label{fig:Pk}
\end{figure}

\begin{figure}
    \centering
    \includegraphics[width=\linewidth]{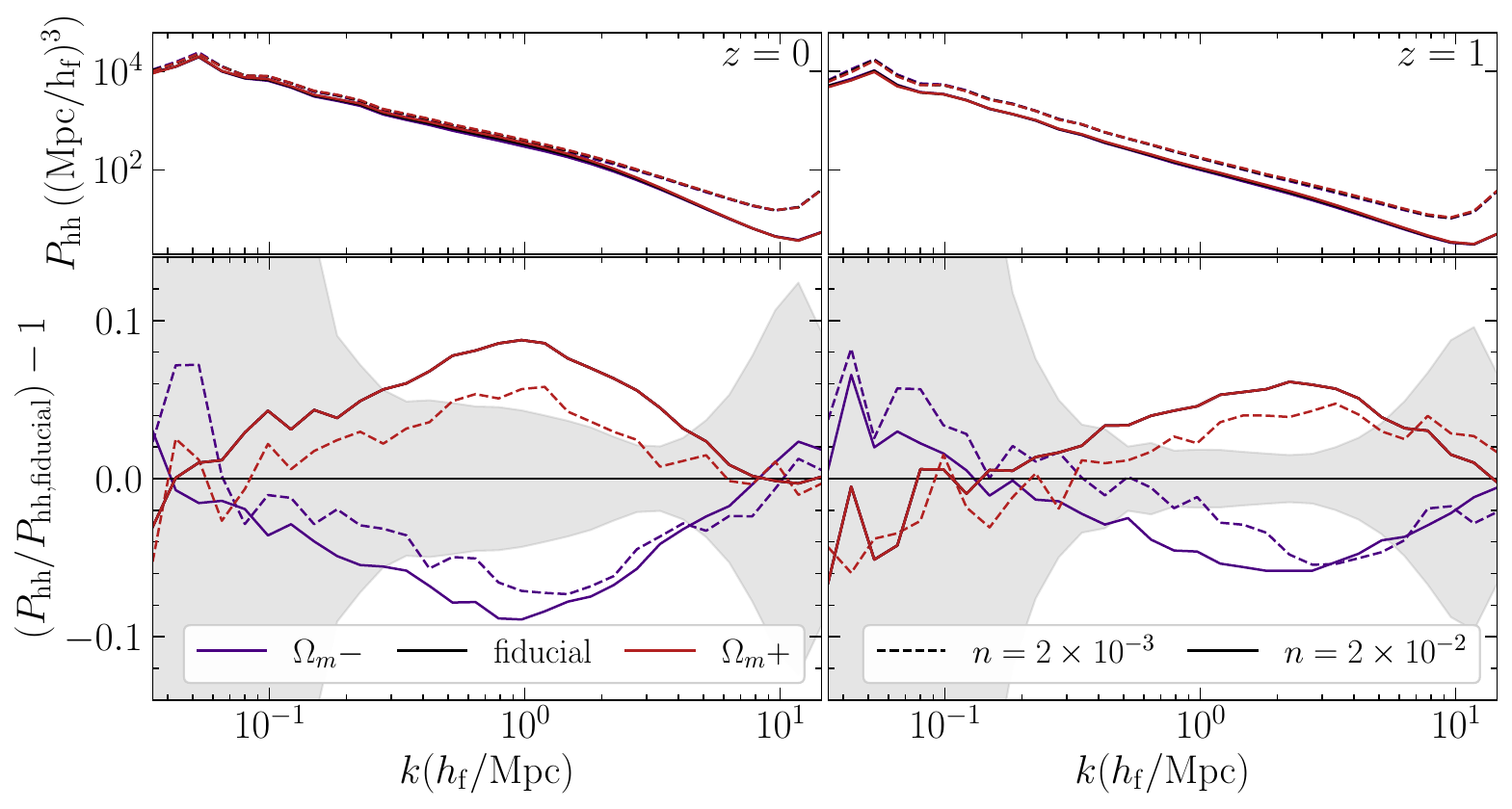}
    \caption{
    Comparison of the halo power spectrum ($P_{\rm{hh}}$) for different tracers, \Om values, and redshifts. The top panels show $P_{\rm{hh}}$ for two tracer populations with different number densities, thresholded on \Vpeak (see text for details). Solid lines correspond to the fiducial cosmology, while dashed (dotted) lines indicate the $\Om+$ ($\Om-$) variations. The bottom panels show the ratio of each $P_{\rm{hh}}$ to its fiducial counterpart, following the same conventions as Fig.~\ref{fig:Pk}. Grey bands show the jackknife errors on the highest number density sample halo power spectrum
    }
    \label{fig:Pk_hh}
\end{figure}

Figure~\ref{fig:Pk} shows the matter auto power spectrum and its response to variations in \Om. The measured power spectra from \Sahyadri\, are in good agreement with expectations from \textsc{halofit} \citep{Takahashi2012} across the range of scales shown, and the expected cosmological dependence is clearly recovered. This demonstrates that the simulations robustly capture both the amplitude and scale dependence of clustering in the quasi-linear and mildly non-linear regime. At $z=0$, deviations from \textsc{halofit} begin to appear for $k\gtrsim  4\,\Mpch$. However, this behaviour is expected, given that \textsc{halofit} was calibrated using lower-resolution simulations and relies on interpolation across cosmological parameter space.  Appendix~\ref{App:fitfuncs} shows a detailed comparison of the \Sahyadri\ matter auto power spectra with \textsc{halofit} expectations. 

Figure \ref{fig:Pk_hh} shows the halo auto power spectra for the 
\Vpeak-selected tracer samples. We focus on the two higher number density samples, as the lowest density sample is dominated by noise. Among these, the highest density sample exhibits the clearest and most robust cosmological trends across cosmic epoch.
For comparison, we estimate the uncertainties using the delete-one jackknife procedure in which the simulation volume is divided into $10\times10$ cuboidal regions in the $x-y$ plane, each spanning the full extent of the box along the $z$-direction. The halo power spectrum is calculated by leaving one region out at a time, and accounting for the reduced volume effect while calculating the density contrast. The resulting jackknife errors from these 100 realizations for the highest-density sample are shown as a shaded grey band in the figure. These errors are  for illustrative purpose only and are not intended for quantitative inference. We see from the Figure that, owing to its high resolution, \Sahyadri\ probes deeply non-linear, highly non-Gaussian scales up to $k\gtrsim 10\Mpch$.

\begin{figure}
    \centering
    \includegraphics[width=\linewidth]{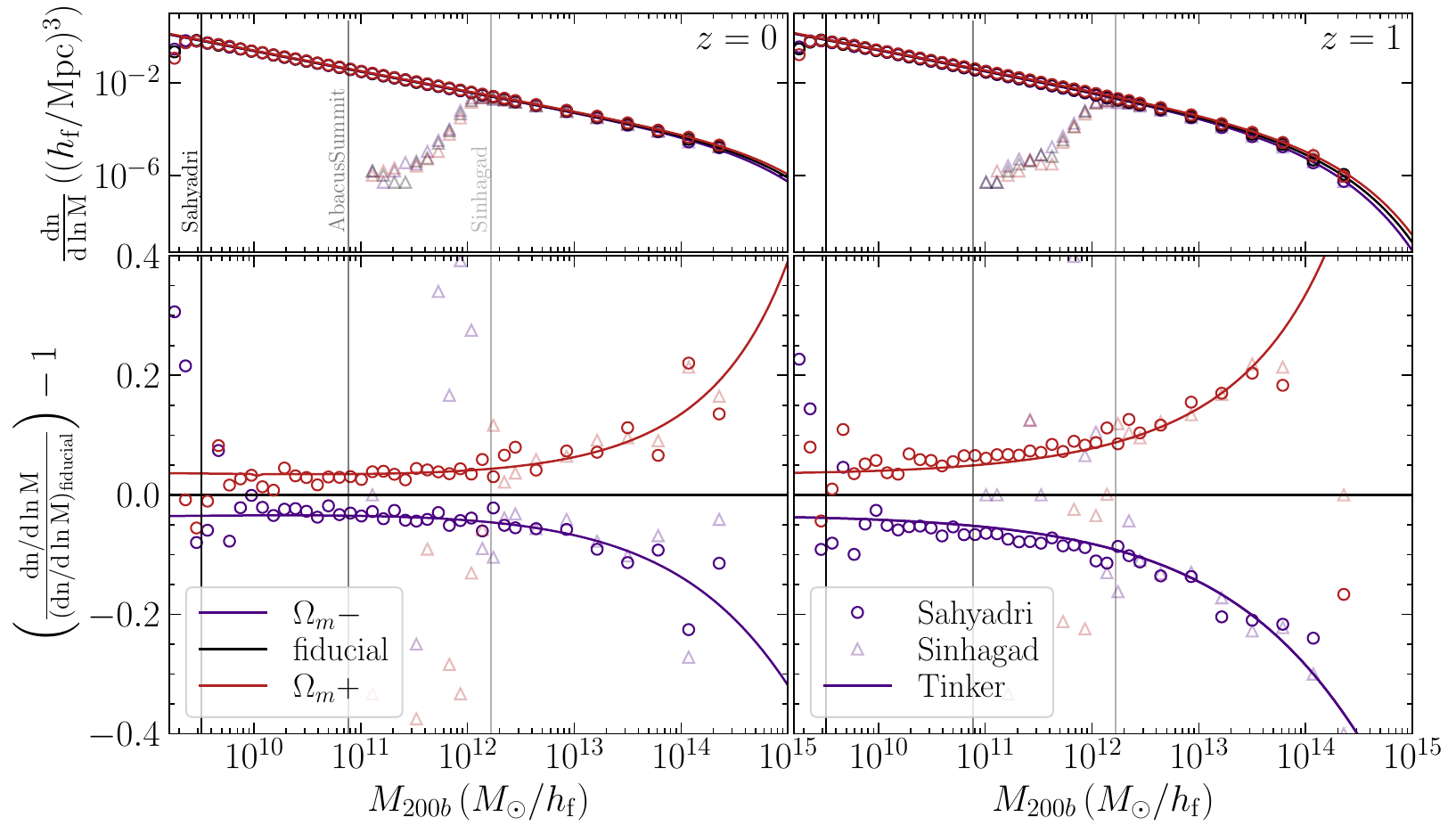}
    \caption{Variation of the mass function as a function of \Om. The arrangement of colors and panels follows Fig.~\ref{fig:Pk}. Solid lines show the Tinker mass function expectation, while the grey vertical lines mark the 40-particle limit for the fiducial \Sahyadri, AbacusSummit and \Sinhagad\, simulation. 
    }
    \label{fig:mf}
\end{figure}

Figure \ref{fig:mf} shows the variation of the halo mass function with \Om, for mass-selected samples. The measured mass functions are in good agreement with the Tinker \citep{Tinker+2008} prediction over the resolved mass range, and the impact of cosmological variations is clearly detected. The extended low-mass reach of \Sahyadri\ allows these trends to be explored well below the mass thresholds accessible to existing simulation suites. Appendix~\ref{App:fitfuncs} shows a detailed comparison with Tinker expectation.

\begin{figure}
    \centering
    \includegraphics[width=\linewidth]{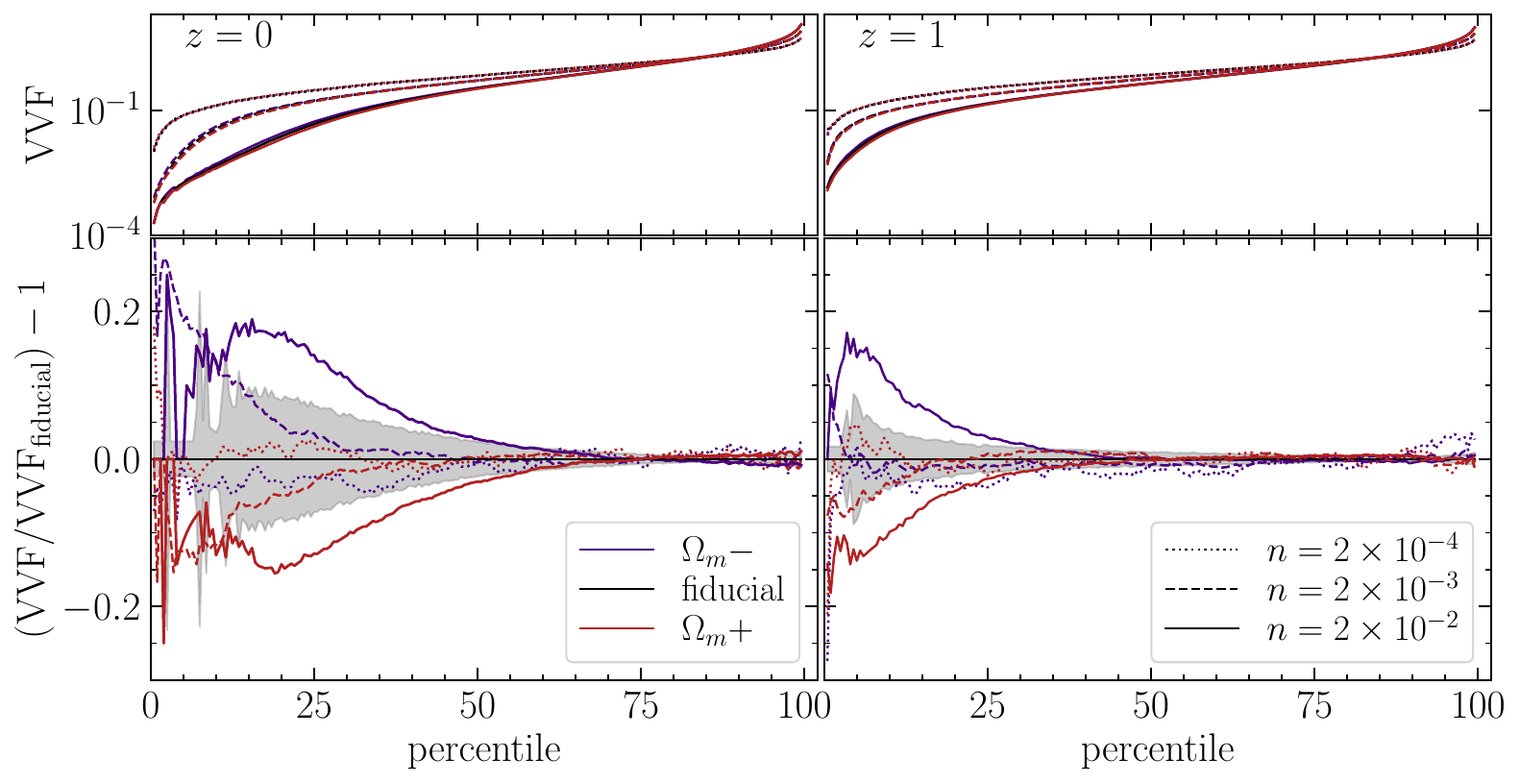}
    \caption{Same as Figure~\ref{fig:Pk_hh}, but for VVF. Here, we show all the three number density tracers considered in this paper. Grey bands show the jackknife errors on the higher number density VVF. The results suggest that the VVF, particularly for the higher density tracers, has significant potential to constrain \Om.}
    \label{fig:vvf_100}
\end{figure}

We now turn to statistics that probe the geometry of the cosmic web beyond standard number counts or 2-point clustering measures. Figure \ref{fig:vvf_100} shows the cosmological dependence of the Voronoi volume function (VVF) for three \Vpeak-selected tracer samples. The VVF, introduced in \citep{VVF2020}, is defined as the inverted cumulative distribution function of Voronoi cell volumes \citep{Voronoi1908} and is sensitive to the full hierarchy of higher-order correlations encoded in the tracer distribution.

The improved mass resolution of \Sahyadri\ allows us to compute the VVF for tracer samples with substantially higher number densities than previously explored. The physical length scales corresponding to the VVF percentiles are shown in Appendix \ref{app:vvf-scale}. For the highest-density sample, the statistic probes deeply non-linear scales down to $\sim 0.1\rm{Mpc}$.
The delete-one jackknife errors for the highest density tracer sample are shown as the grey band in Figure \ref{fig:vvf_100}. The VVF exhibits a clear and systematic response to variations in \Om, with the strongest sensitivity observed for the highest density tracer sample, where the cosmological signal significantly exceeds the estimated uncertainties over a wide range of percentiles. As shown in \citep{dhawalikar2025stabilizingsimulationbasedcosmologicalfisher}, lower density samples suffer from increased noise, requiring stabilization techniques to identify subsets of percentiles that provide robust Fisher constraints. In contrast, the high-density samples accessible with \Sahyadri\ display smooth and coherent variations of the VVF with cosmology, enabling substantially cleaner measurements and improved constraining power.

\begin{figure}
    \centering
    \includegraphics[width=\linewidth]{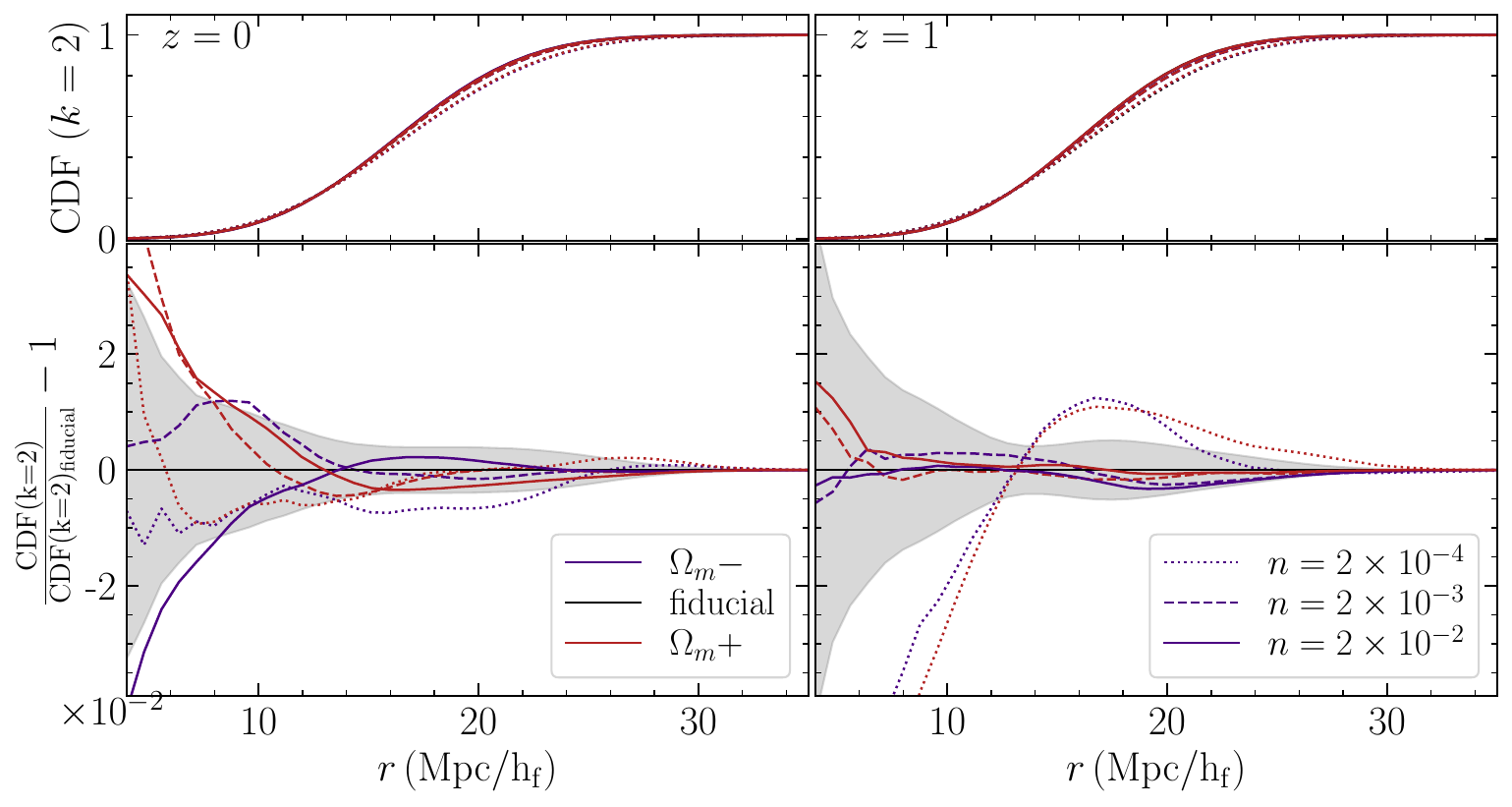}
    \caption{Same as Figure~\ref{fig:vvf_100}, but for the $k$NN CDF at $k=2$.}
    \label{fig:knn}
\end{figure}

A related set of summary statistics that can be used to quantify differences in higher-order halo clustering across cosmologies and samples are the $k$-Nearest Neighbor ($k$NN) distributions. Introduced in \cite{Banerjee:2020}, these statistics are also formally sensitive to all connected $N$-point functions of the underlying clustering \cite{Banerjee:2020, Banerjee:2021, Banerjee:2022} and can be directly related to the geometric features of the cosmic web~\cite{Gangopadhyay:2025}. The $k$-Nearest Neighbor Cumulative Distribution functions ($k$NN-CDFs) are computed by first populating the simulation box with a random, volume filling set of query points, and then creating a sorted list of distances from these query points to the $k$-th nearest neighbor data (halo) points. Using a $k$-d tree structure, these calculations can be sped up to $\mathcal O (N \log N)$ scaling. In Figure \ref{fig:knn}, we plot the results for $k=2$ at redshifts $z=0$ \emph{(left panels)} and $z=1$ \emph{(right panels)} over the radial range $[5,35]\,h^{-1}$Mpc. The $k$NN-CDFs are sensitive to the number density; therefore, to plot them on the same radial scale, we have split halos from each sample into sets with fixed number densities $1\times 10^{-4} (\hMpc)^3$ before performing the calculations. The bottom panels in each column plot the residuals with respect to measurements of the $k$NN-CDFs of that halo sample in the fiducial cosmology. At $z=0$, the sample with an inherent number density of $n=2\times 10^{-2}$ show a clear trend with $\Omega_m$ that is statistically significant. Interestingly, for the other samples, the tradeoff between the underlying clustering of the matter field (controlled by $\Omega_m$), and the change in the bias values of the tracers at a fixed inherent number density nearly cancels each other out over a certain range of scales in the plot. The $k$NN-CDFs are sensitive to both of these effects~\cite{Banerjee:2021cmi}, and  by varying the number density, can be used to break potential degeneracies.

The VVF and $k$NN results above highlight the potential of higher-order, cosmic web–based statistics to extract cosmological information from the quasi-linear and non-linear regimes enabled by the resolution of \Sahyadri.

\begin{figure}
    \centering
    \includegraphics[width=\linewidth]{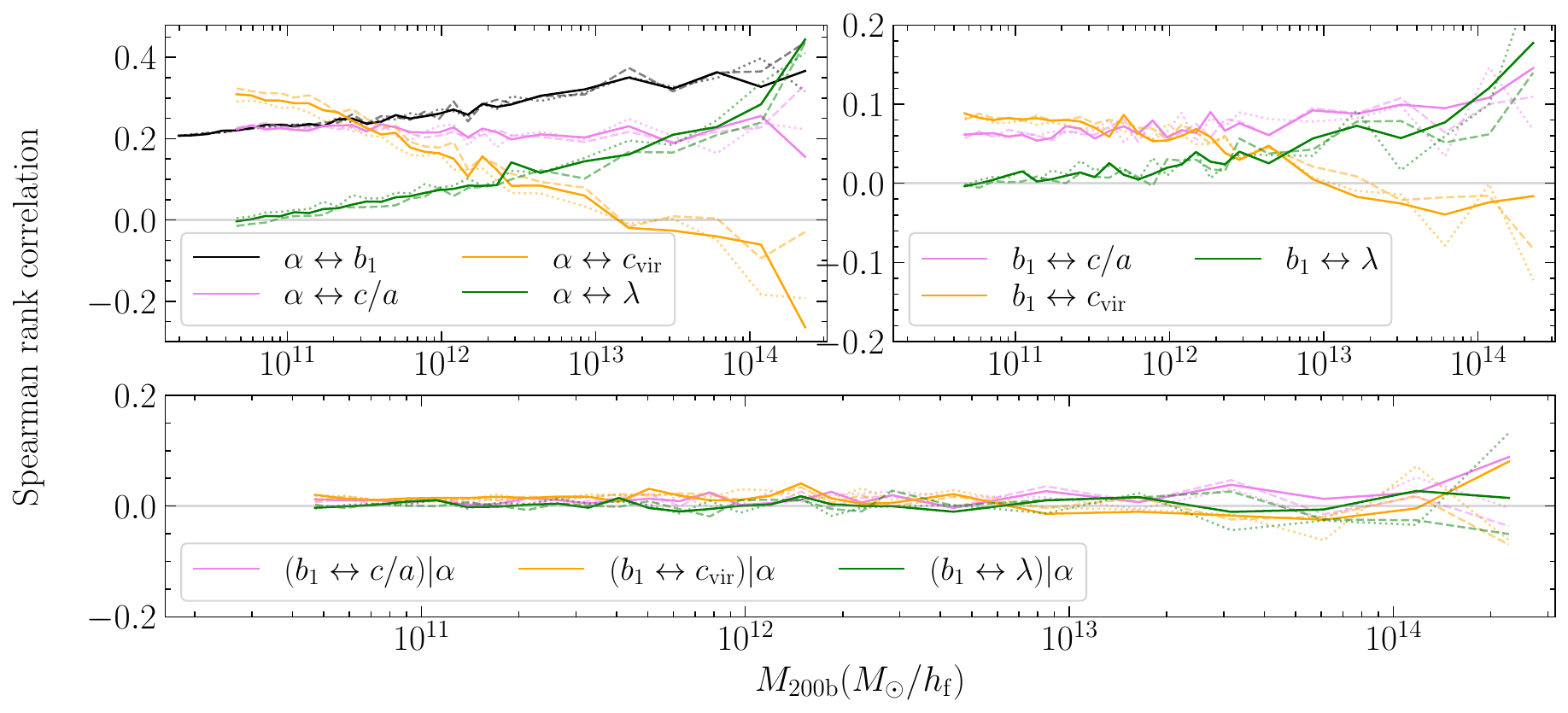}
    \caption{Spearman rank correlations between halo environment and internal properties, as a function of halo mass. Solid lines correspond to the fiducial cosmology, while dashed (dotted) lines indicate the $\Om+$ ($\Om-$) variations. With Sahyadri, these correlations can now be measured for halos a factor of $\sim18–45$ smaller than previously accessible (see text for details).
    }
    \label{fig:correlations}
\end{figure}

\begin{figure}
    \centering
    \includegraphics[width=\linewidth]{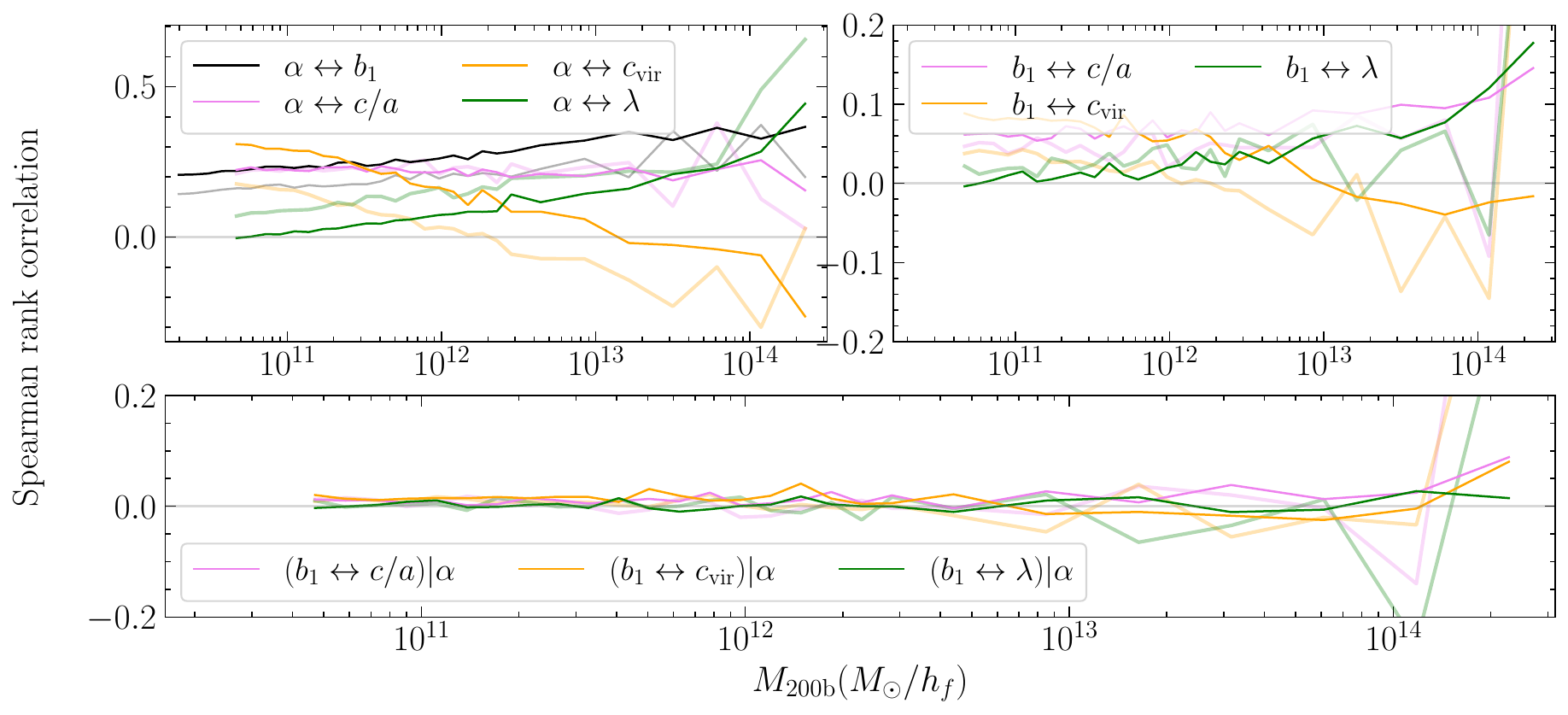}
    \caption{Redshift dependence of correlations for the fiducial cosmology. The panels are the same as in Fig~\ref{fig:correlations}, with dark (pale) curves representing $z=0 \, (1)$.}
    \label{fig:correlations_z}
\end{figure}

Finally, following the approach of \citep{Ramakrishnan+2019}, Figures \ref{fig:correlations} and \ref{fig:correlations_z} examine the correlations between halo environment and internal halo properties in our mass-selected samples, and their dependence on cosmology and redshift. We consider three internal properties: the halo concentration $c_{\rm vir}$, the mass ellipsoid axis ratio $c/a$, which characterizes halo shape, and the dimensionless spin parameter $\lambda$, which quantifies halo angular momentum. To ensure robust measurements of these quantities, we restrict the analysis to halos resolved with at least 500 particles. For the $\Om +$ cosmology, which has the largest particle mass, this corresponds to a minimum halo mass of $4.2\times10^{10}\,\Msun/h$; this mass threshold is adopted uniformly throughout the analysis.

Halo environment is characterized using the tidal anisotropy $\alpha$ (equation~\ref{eq:alpha-def}) and the halo-by-halo large-scale bias $b_1$, both evaluated on a $1024^3$ grid (see section~\ref{subsec:halos_vahc} for details). Following Appendix A1 of \citep{phs18}, we further require that halos be resolved by at least eight grid cells within a sphere of radius $4R_{\rm 200b}$. For the $\Om +$ cosmology, this corresponds to a mass threshold of $1.7\times10^{10}\,\Msun /h$, which we again apply consistently across all cosmologies. Correlations are quantified using Spearman’s rank correlation coefficient.
Figures \ref{fig:correlations} and \ref{fig:correlations_z} present the resulting correlations \emph{(top panels)}, together with the conditional correlation coefficients with distributions conditioned on $\alpha$ \emph{(bottom panels)}. The qualitative trends are consistent with those reported by \citep{Ramakrishnan+2019}, and with related studies demonstrating that correlations between halo environment and internal halo properties encode much of the observed assembly bias, with tidal anisotropy $\alpha$ playing a key role, extending across mass, redshift, and cosmology \citep[e.g.][]{Ramakrishnan+2025, RamakrishnanVelmani2022,RPS2021}.
However, owing to the substantially improved mass resolution of \Sahyadri, we are able to extend this analysis to halos that are smaller by factors of $\sim45$ for the $b_1\leftrightarrow\alpha$ correlation and $\sim18$ for correlations involving internal halo properties. 
We see that $\alpha$ is positively correlated with $b_1$ and $\lambda$, with the correlation strength increasing with halo mass. The correlation between $\alpha$ and $c_{\rm{vir}}$ is positive at lower masses, and crosses over to becomes negative for $M_{\rm{200b}} \gtrsim 10^{13}\Msun /h$ \cite{phs18,Ramakrishnan+2019}. The correlation with $c/a$ remains positive and comparatively independent of mass. The \emph{top right panel} shows that the correlation of internal properties with $b_1$ is much weaker than that with the tidal anisotropy $\alpha$, while the \emph{bottom panel} highlights the central result of \citep{Ramakrishnan+2019}:  conditional correlation coefficients are significantly smaller relative to their unconditional counterparts, indicating that $\alpha$ is largely responsible for all of these assembly bias trends. No strong trends are seen with variations in \Om.

Overall, the enhanced resolution and tracer density of \Sahyadri\, increase the reach of standard clustering statistics and, at the same time, allow reliable measurements of higher-order, environment-sensitive statistics that probe the cosmic web in the quasi-linear and non-linear regimes.

\section{Summary and Future Outlook}
\label{sec:summary}

In this paper, we have presented \Sahyadri, a new suite of high-resolution $N$-body simulations specifically designed to bridge the gap between the mass resolution required for modeling low-redshift spectroscopic surveys and the cosmological parameter coverage needed for precision inference. The key specifications and results of this work can be summarized as follows:

\Sahyadri\ simulations evolve $2048^3$ dark matter particles in periodic boxes of side length $200\,h^{-1}\,{\rm Mpc}$, achieving a particle mass of $m_{\rm p} = 8.1 \times 10^{7}\,h^{-1}\,M_{\odot}$ and resolving halos down to $M_{\rm min} = 3.2 \times 10^{9}\,h^{-1}\,M_{\odot}$ (40 particles). This represents a factor of $\sim$25 improvement in mass resolution compared to AbacusSummit \citep{Maksimova2021} and enables comprehensive modeling of $\sim$40\% more DESI BGS galaxies at $z < 0.15$ than previously possible. The suite systematically varies six cosmological parameters ($\Omega_{\rm m}$, $h$, $n_{\rm s}$, $A_{\rm s}$,  $w_{\parallel}$, $\Omega_{\rm k}$) around Planck 2018 values using seed-matched initial conditions to enable derivative calculations for Fisher matrix analyses.

We employ \textsc{gadget-4} with 2LPT initial conditions at $z=49$, producing 101 snapshots uniformly spaced in scale factor between $z=12$ and $z=0$. Our custom particle data compression scheme reduces storage requirements by a factor of $\sim$3 while maintaining clustering accuracy at the sub-percent level, achieving a total data footprint of $\sim$117 TB for the complete suite (including FITS-compressed halo and value-added catalogs). The simulations utilize seed-matched initial conditions across cosmological variations to suppress sample variance and isolate cosmological effects.

We identify halos using the \textsc{Rockstar} phase-space halo finder \citep{Rockstar1023} and implement quality control criteria to ensure robust mass estimates and remove spurious objects.
 Our convergence tests demonstrate that finite volume effects on the linear halo bias become negligible ($\lesssim 1\%$) for halos with $M \gtrsim 10^{12}\,h^{-1}\,M_{\odot}$ (Appendix~\ref{app:b1-voldep}). The measured matter power spectra and halo mass functions show excellent agreement with Halofit and Tinker predictions, respectively, validating our numerical methods and force resolution choices (Appendix~\ref{App:fitfuncs}).

We have showcased \Sahyadri's capabilities through:
\begin{itemize}
\item Visualizations of cosmic web evolution from $z=5.6$ to $z=0$, highlighting the emergence of filaments, nodes, and the increasing density contrast.
\item Direct comparisons with AbacusSummit-resolution halos, demonstrating the enhanced sampling of filaments, cluster outskirts, and underdense regions enabled by \Sahyadri's improved mass resolution.
\item Voronoi-based density field reconstructions estimates showing progressively better resolution of cosmic web structure with increasing tracer number density.
\item Measurements of the cosmological dependence of standard clustering, abundance, and environment statistics — including the matter and halo auto power spectra, halo mass function, and correlations between halo environment (tidal anisotropy and halo bias) and internal halo properties (concentration, spin, and shape) as a function of mass at $z=0$ and $z=1$. While these trends are consistent with earlier work, \Sahyadri\ extends such analyses to significantly lower halo masses than previously accessible, with the matter power spectrum included mainly as a consistency check. 
\item First measurements of the Voronoi volume function (VVF) and $k^{\rm th}$-nearest neighbour ($k$NN) statistics  for high-density halo samples ($n \sim 10^{-2}\,{\rm Mpc}^{-3}$), revealing significant sensitivity to $\Omega_{\rm m}$ variations.
\end{itemize}

The enhanced resolution of \Sahyadri\ is particularly well-suited for addressing several pressing questions in contemporary cosmology. The low-redshift Universe ($z < 0.5$) is the regime where surveys achieve their highest number densities and where potential solutions to the Hubble tension and hints of evolving dark energy may reside. Traditional 2-point statistics in this regime have begun to approach cosmic variance limits given available volumes. \Sahyadri\ enables the exploration of higher-order statistics and smaller-scale clustering in a controlled framework with cosmological parameter variations, opening new avenues for extracting cosmological information from the non-linear regime.

The \Sahyadri\ simulations enable diverse applications including halo occupation distribution modeling for DESI BGS and 4MOST, studies of environmental quenching and AGN activity in low-mass halos, and validation of analytical models bridging quasi-linear and non-linear scales. The enhanced resolution facilitates Fisher matrix forecasts for beyond-two-point statistics (VVF, $k$NN, Minkowski functionals, etc.), improved estimates of fiber collision effects in spectroscopic surveys through better sampling of faint galaxies, and construction of training datasets for machine learning-based galaxy-halo connection studies in the non-linear regime.

As of this publication, we have completed the fiducial cosmology simulation along with the full parameter variations for \Om, $h$, \ns, and \As. The remaining parameter variations ($w_{\parallel}$, $\Omega_{\rm k}$) are in progress. We will also release higher-level data products, including detailed merger trees, pre-computed summary statistics (power spectra, correlation functions, VVF, $k$NN statistics, etc.) across all cosmologies, mock galaxy catalogs using empirical galaxy-halo connection models calibrated to DESI, SDSS, and GAMA observations, and cut-sky catalogs for direct comparison with survey geometries.

All \Sahyadri\ data products—including particle snapshots, halo catalogs, compression tools, and analysis scripts will be made publicly available to the community through a dedicated web portal. We anticipate that \Sahyadri\ will become a valuable resource for the broader community working on low-redshift large-scale structure, galaxy evolution, and non-linear cosmology, complementing existing simulation suites and enabling new science in the transition regime between quasi-linear theory and full hydrodynamical modeling.

\section*{Data availability}
Our pipeline for performing the simulations, generating halo and environment catalogs and other post-processing products, along with data compression tools, is currently publicly available at \url{https://github.com/a-paranjape/sahyadri-sandbox}. Downsampled snapshots, as well as the complete halo and value-added catalogs at selected redshifts, are publicly available through the dedicated website at  \url{https://sahyadri.tifr.res.in/index.html} . Additional products not included in the public release may be requested by contacting the authors.

We encourage the community to explore potential applications of \Sahyadri\ and welcome collaborations. Feedback on data products, analysis tools, and scientific priorities for future releases is highly appreciated. 

\acknowledgments
It is a pleasure to acknowledge many valuable, free-flowing discussions with the participants of the Pune-Mumbai Cosmology and Astroparticle Physics (PM-CAP) series of meetings,\footnote{\url{https://www.tifr.res.in/~shadab.alam/PM_CAP_meeting/}} which is where the idea for the \Sahyadri\ suite originated. The name \Sahyadri\ derives from a well-known mountain range between Pune and Mumbai, while the pilot \Sinhagad\ suite is named after a popular peak in this range. AP thanks Emiliano Sefusatti for valuable discussions.
The research of AP is supported by the Associates Scheme of ICTP, Trieste. AB's work was partially supported by grants SRG/2023/000378 and ANRF/ARGM/2025/000301/TS from the Anusandhan National Research Foundation (ANRF) India. SA was supported by the Department of Atomic Energy, Government of India, under Project Identification Number RTI-4012.
This work made extensive use of the open source computing packages NumPy \citep{vanderwalt-numpy},\footnote{\url{http://www.numpy.org}} SciPy \citep{scipy},\footnote{\url{http://www.scipy.org}} Pandas \citep{mckinney-proc-scipy-2010, reback2020pandas},\footnote{\url{https://pandas.pydata.org}} Matplotlib \citep{hunter07_matplotlib},\footnote{\url{https://matplotlib.org/}} and Jupyter Notebook.\footnote{\url{https://jupyter.org}} 
We gratefully acknowledge HPC facilities at IUCAA, Pune and TIFR, Mumbai.

\bibliography{references}

\appendix
\section{Comparison with literature fitting functions}
\label{App:fitfuncs}
\begin{figure}[h!]
    \centering
    \includegraphics[width=\linewidth]{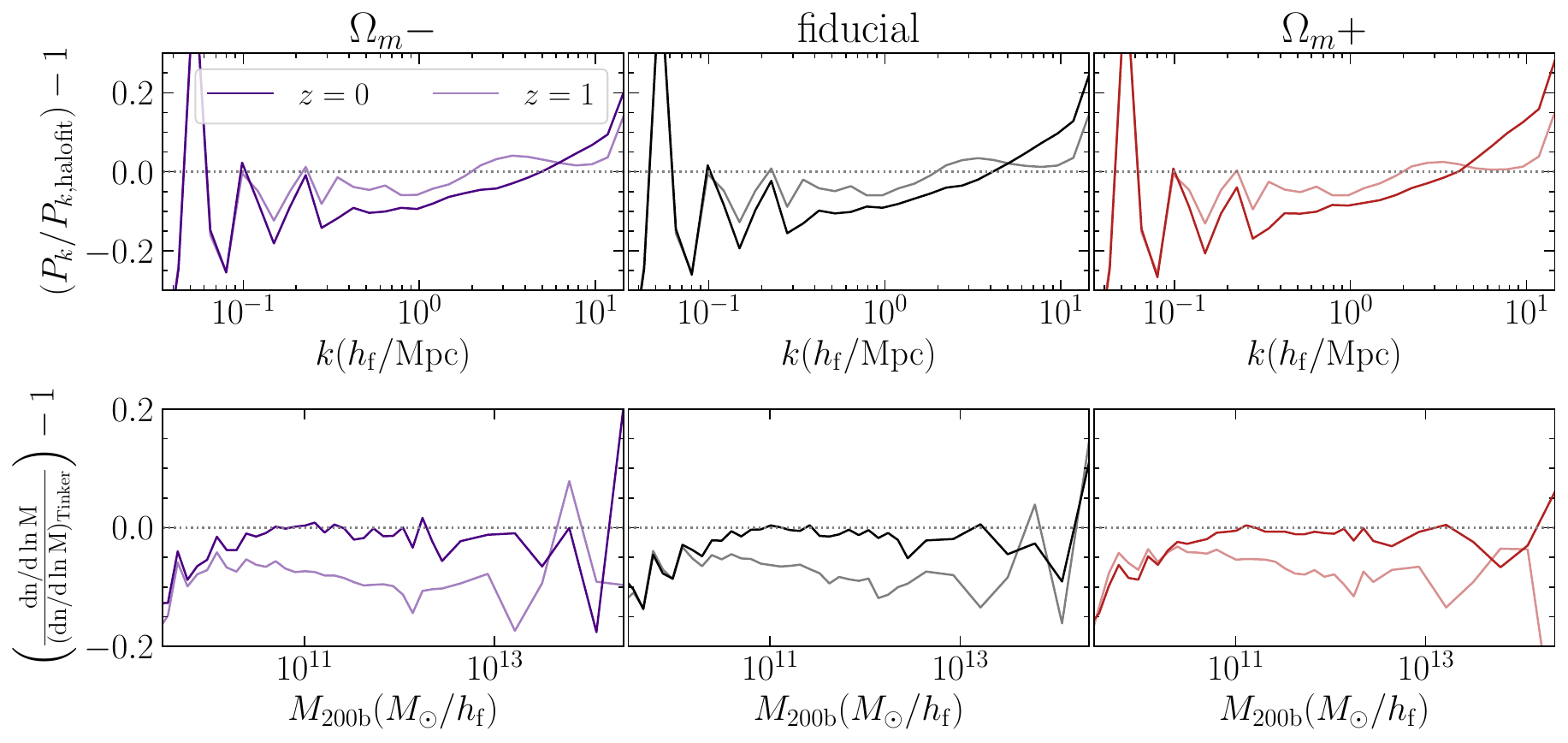}
    \caption{Comparison of the simulation results with expectations from literature. Top (bottom) panels show the ratios of simulation matter power spectrum (halo mass function) with the corresponding halofit (Tinker) expectations.}
    \label{fig:fitfuncs}
\end{figure}

As a consistency check, we compare the matter power spectrum and halo mass function measured from our simulations with widely used fitting functions from the literature, namely \textsc{halofit} \citep{Takahashi2012} for the matter power spectrum and the Tinker mass function \citep{Tinker+2008}. Figure~\ref{fig:fitfuncs} shows these comparisons, with the top panels displaying the matter power spectrum and the bottom panels showing the halo mass function. The three columns correspond to the three cosmologies considered in this work, and results are shown at $z=0$ and $z=1$.

The halo mass function shows excellent agreement with the Tinker fit at $z=0$, with deviations typically below $\lesssim 5\%$ across the resolved mass range. At $z=1$, the level of agreement is slightly weaker, with deviations reaching up to $\sim 15\%$, which is consistent with expectations given the increased uncertainty in fitting functions at higher redshifts.

For the matter power spectrum, we find that at $z=0$ the measured power is systematically lower than \textsc{halofit} on large scales and higher on small scales, with deviations reaching up to $\sim 20\%$. At $z=1$, the discrepancies are reduced, remaining at the level of $\lesssim 10\%$ over the range of scales shown. These trends are consistent with previous studies, e.g. see figure~3 in  \citep{Klypin2016} for halo mass function and figure~1 in \citep{Daalen+2011}.

\section{Length scales in VVF}
\label{app:vvf-scale}
\begin{figure}[h!]
    \centering
    \includegraphics[width=0.9\linewidth]{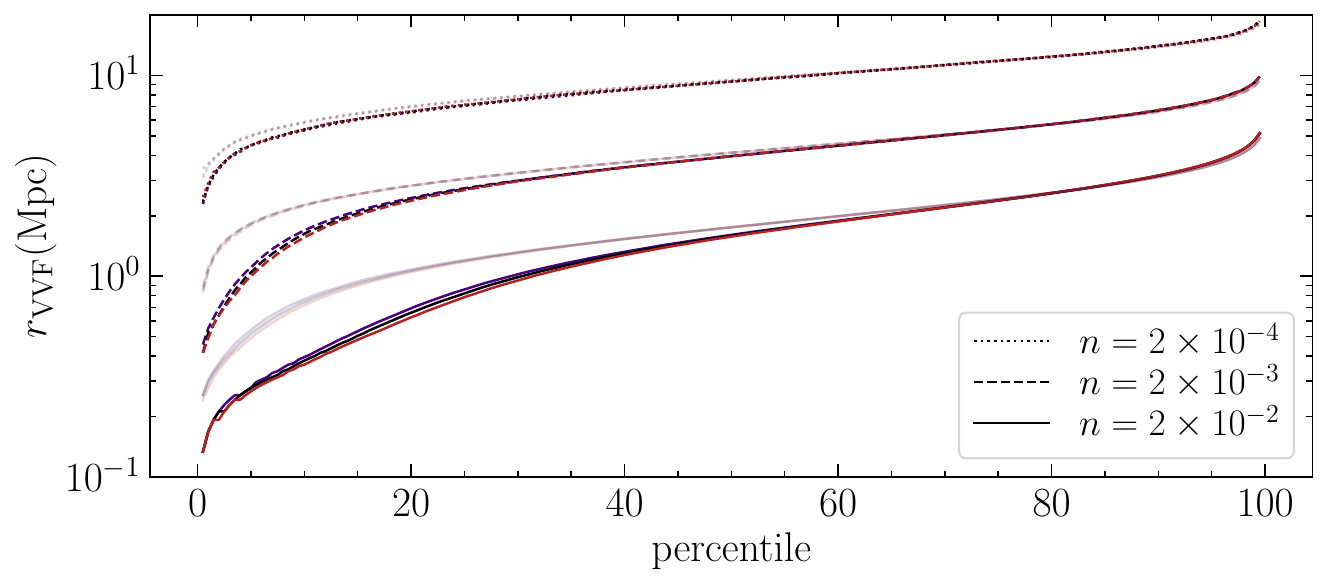}
    \caption{Effective length scales corresponding to VVF percentiles for all tracer number densities, \Om\, variations, and redshifts considered in this paper. The effective scale $r(p)$ is obtained by associating each Voronoi cell volume at percentile $p$ with the radius of a sphere of equal volume. Dark (pale) curves correspond to $z=0$ ($z=1$). The color and linestyle scheme follows that of Figure~\ref{fig:vvf_100}.
    }
    \label{fig:rVVF}
\end{figure}

The VVF is a relatively new statistic, and this work presents its first application to high-density tracer samples. It is therefore useful to identify the physical length scales probed by the VVF. Since the statistic is normalized by the mean tracer number density, it does not explicitly retain absolute spatial scales. These can, however, be recovered by associating VVF percentiles with effective length scales.\\
The VVF at percentile $p$ is defined as
\begin{align}
\mathrm{VVF}(p) &= \frac{V(p)}{\langle V \rangle} = nV(p),
\end{align}
where $V(p)$ is the Voronoi cell volume at percentile $p$, $\langle V \rangle$ is the mean cell volume, and $n$ is the tracer number density. Defining an effective spherical radius as:
\begin{align}
\frac{4}{3}\pi r^3(p) &= V(p), \
r(p) = \left(\frac{3}{4\pi}V(p)\right)^{1/3},
\end{align}
gives a characteristic length scale corresponding to each VVF percentile.

Figure~\ref{fig:rVVF} shows these length scales for all tracer number densities, cosmological variations, and for both redshifts considered in this paper. Dark (pale) curves correspond to $z=0$ ($z=1$), with the color and linestyle scheme matching Figure~\ref{fig:vvf_100}. The VVF probes scales ranging from $\sim 0.1,\mathrm{Mpc}$ to $\sim 20,\mathrm{Mpc}$, with the range depending on tracer number density and redshift. Higher number densities probe systematically smaller scales, as expected. We also find a clear redshift dependence of the effective length scales associated with the VVF percentiles. At fixed tracer number density, percentiles below $\sim 80$ correspond to systematically larger effective radii, and hence larger absolute Voronoi cell volumes, at $z=1$ compared to $z=0$; while the highest percentiles are largely unchanged. This behaviour reflects the less clustered nature of the tracer distribution at higher redshift, which leads to a narrower distribution of Voronoi volumes with fewer extreme small cells associated with dense environments.

Comparison with Figure~\ref{fig:vvf_100} shows that, for the highest-density tracer sample, the strongest response to changes in $\Om$ occurs at scales $\lesssim 1,\mathrm{Mpc}$. This indicates that the VVF is sensitive to information from the non-linear regime.

\section{Volume dependence of $b_1$}
\label{app:b1-voldep}

\begin{table*}[h!]
\centering
\begin{tabular}{lcccc}
\hline\hline
Simulation & Cosmology & $L_{\rm box}$ & $N_{\rm part}$  \\
 &  & $[h^{-1}\,{\rm Mpc}]$ &  \\
\hline
A & C1 & 150 & $1024^3$\\
B & Planck18 & 200 & $256^3$\\
C & C1 & 300 & $1024^3$\\
D & Planck18 & 400 & $512^3$\\
E & C1 & 600 & $1024^3$\\

\hline
\end{tabular}
\caption{Summary of simulations used to study the volume dependence of halo-by-halo bias $b_1$. The default \Sahyadri\, cosmology is Planck18, while C1 corresponds $\Om=0.276, \Ob=0.045, h=0.7, \ns=0.961, \sigma_8=0.811$.} 
\label{tab:diffvol_sims}
\end{table*}

\begin{figure}
    \centering
    \includegraphics[width=\linewidth]{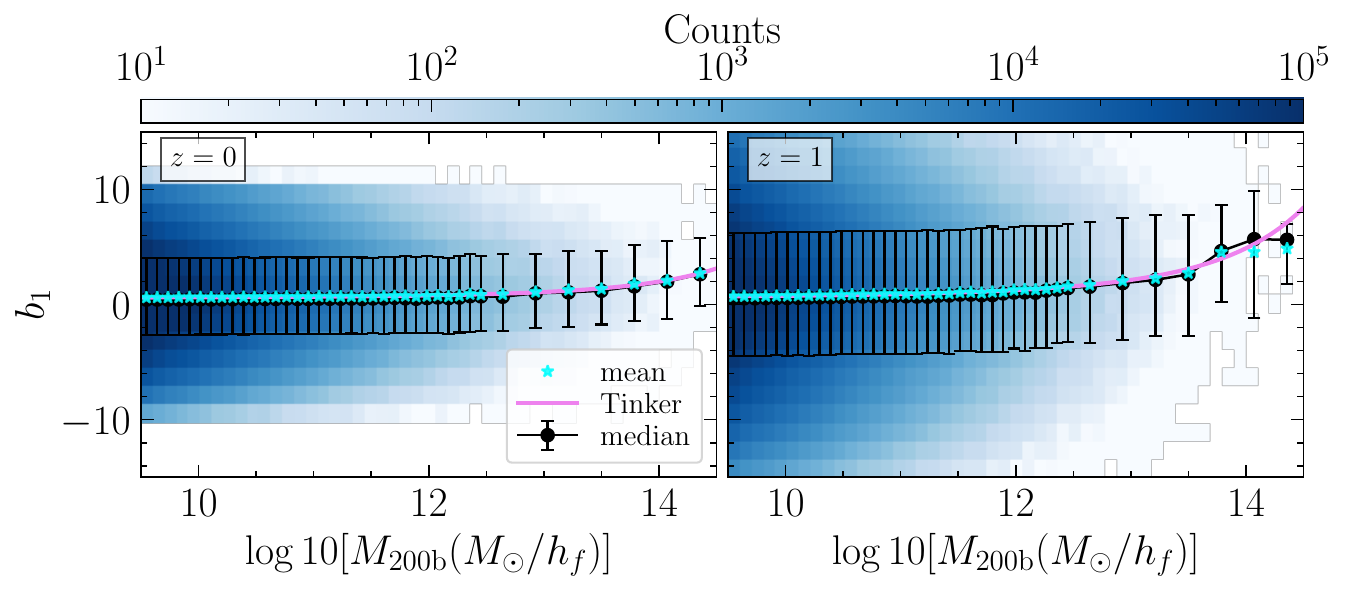}
    \caption{Distribution of halo-by-halo bias $b_1$ as a function of halo mass at $z=0$ (left) and $z=1$ (right). The background shows the 2D histogram of halo counts in the $(M_{\rm 200},b_1)$ plane.
    Black points with error bars indicate the median and the $16$–$84$ percentile range in each mass bin. Cyan stars denote the mean bias, while the violet curve shows the Tinker prediction. }
    \label{fig:b1_2d}
\end{figure}

\begin{figure}
    \centering
    \includegraphics[width=\linewidth]{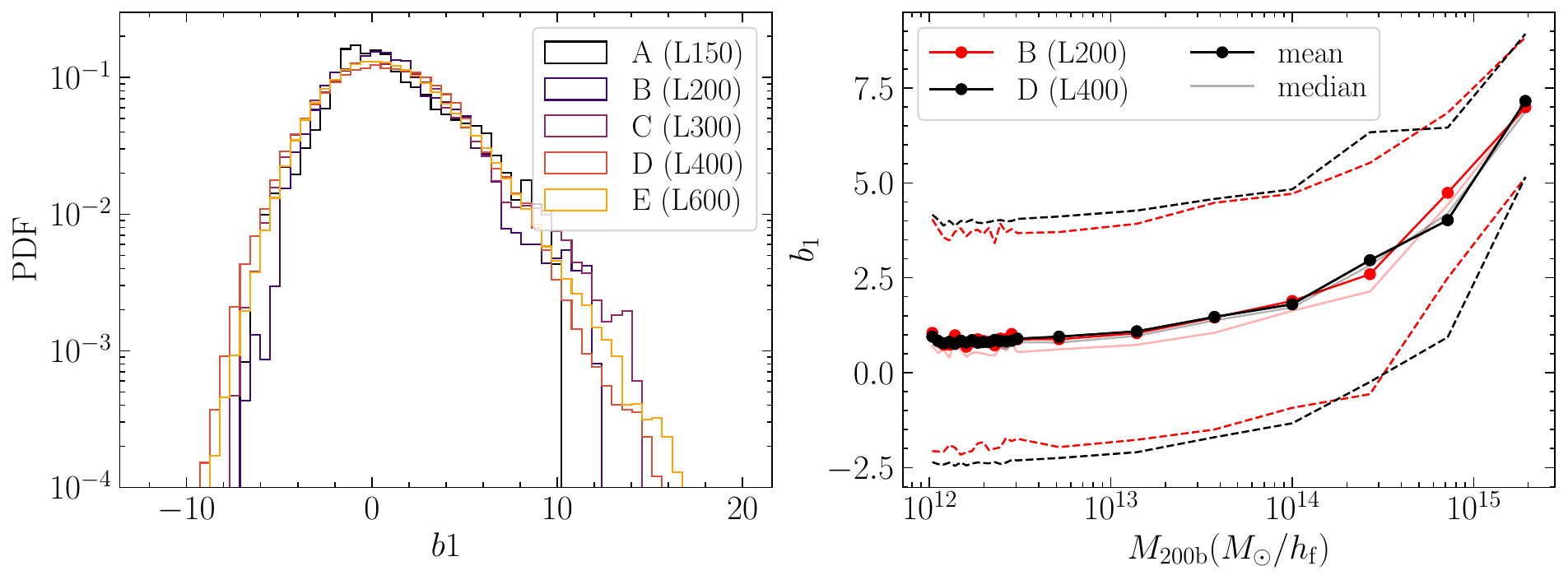}
    \caption{\emph{Left panel:} Distribution of $b_1$ for five simulations with varying box sizes (box size in brackets, in units of $\mathrm{Mpc}/h$). Smaller volumes exhibit truncated tails in $b_1$, indicating the absence of extreme environments. \emph{Right panel:}Comparison of $b_1$ as a function of mass between two simulations with identical cosmology and resolution, but different volumes. Dark (pale) curves show the mean (median) bias in each mass bin, while dashed lines indicate the $16$–$84$ percentile range. 
    }
    \label{fig:b1_vol}
\end{figure}

The halo-by-halo bias $b_1$ plays an important role in the environment-based analyses presented in this paper. Since $b_1$ is sensitive to the surrounding large-scale density field, its measurement can be affected by finite-volume effects, particularly in the tails of the bias distribution. In this Appendix, we examine how its distribution and mass dependence varies with simulation volume.

Figure~\ref{fig:b1_2d} shows the distribution of $b_1$ in bins of halo mass for $z=0, 1$, using halos cleaned with the QE criterion and restricted to parent halos in the default \Sahyadri\, simulation. The mean values of $b_1$ agree well with the Tinker expectation \citep{Tinker+2010}. However, we see a sharp truncation at very low and high values of $b_1$, at $z=0$. This is because the bias, being influenced by the large-scale environment, may be sensitive to finite-volume effects which lead to missing long-wavelength modes in the simulation volume.
 
To study this, we consider five different simulations spanning a range of volumes. Simulations B, D have the same cosmology as our fiducial one, whereas simulations A, C, E have a slightly different cosmology as summarized in Table~\ref{tab:diffvol_sims}. Simulations A, C and E are the same as those used by \citep{NPaul+2019}, whereas B is the default, low-resolution \Sinhagad\ simulation. We clean the halos based on the QE criteion, and retain only parent halos with $M_{\rm 200}\geq 10^{12} \Msun/h$. Figure~\ref{fig:b1_vol} illustrates the volume dependence of $b_1$. The \emph{left panel} shows the distribution of $b_1$ for these simulations. We see that the tails of the distribution are progressively suppressed in lower volumes, reflecting the absence of rare overdense and underdense environments in smaller simulation boxes.

The \emph{right panel} of the Figure shows a comparison between the mass dependence of $b_1$ for the default \Sinhagad\, simulation (labeled B), and a simulation with identical cosmology and mass resolution but eight times larger volume (labeled D). Dark (pale) curves represent the mean (median) value of $b_1$ in each mass bin, while the dashed curves show the $16$th and $84$th percentiles. The median bias shows a systematic dependence on volume, with smaller volume consistently under-estimating the median $b_1$ by $\Delta b\sim0.2$. The spread $\sigma_b$ of $b_1$ distribution is also under-estimated in the smaller simulation box $\Delta\sigma_b\sim0.7$. In contrast, the mean value remains unbiased. 
Overall, we find that finite-volume effects mainly alter the distribution of halo-by-halo bias by truncating its tails, especially at high $b_1$, while leaving the mean mass–bias relation largely unchanged. Because our analysis is based on Spearman rank correlations, these effects will reduce the dynamic range of $b_1$ and may slightly weaken the inferred correlations, without qualitatively changing the trends.

\end{document}